\documentclass[12pt,preprint]{aastex}

\begin{document}

\shortauthors{Hicken et al.}
\shorttitle{Light Curves for 94 SN Ia}

\title{CfA4:  Light Curves for 94 Type Ia Supernovae}

\author{Malcolm Hicken\altaffilmark{1},
 Peter Challis\altaffilmark{1},
 Robert P. Kirshner\altaffilmark{1},
 Armin Rest\altaffilmark{2},
 Claire E. Cramer\altaffilmark{3},
 W. Michael Wood-Vasey\altaffilmark{4},
 Gaspar Bakos\altaffilmark{1,5},
 Perry Berlind\altaffilmark{1},
 Warren R. Brown\altaffilmark{1},
 Nelson Caldwell\altaffilmark{1},
 Mike Calkins\altaffilmark{1},
 Thayne Currie\altaffilmark{6},
 Kathy de Kleer\altaffilmark{7},
 Gil Esquerdo\altaffilmark{8},
 Mark Everett\altaffilmark{8},
 Emilio Falco\altaffilmark{1},
 Jose Fernandez\altaffilmark{1},
 Andrew S. Friedman\altaffilmark{1},
 Ted Groner\altaffilmark{1},
 Joel Hartman\altaffilmark{1,5}, 
 Matthew J. Holman\altaffilmark{1},
 Robert Hutchins\altaffilmark{1},
 Sonia Keys\altaffilmark{1},
 David Kipping\altaffilmark{1},
 Dave Latham\altaffilmark{1},
 George H. Marion\altaffilmark{1},
 Gautham Narayan\altaffilmark{1},
 Michael Pahre\altaffilmark{1},
 Andras Pal\altaffilmark{1},
 Wayne Peters\altaffilmark{1},
 Gopakumar Perumpilly\altaffilmark{9},
 Ben Ripman\altaffilmark{1},
 Brigitta Sipocz\altaffilmark{1},
 Andrew Szentgyorgyi\altaffilmark{1},
 Sumin Tang\altaffilmark{1},
 Manuel A. P. Torres\altaffilmark{1},
 Amali Vaz\altaffilmark{10},
 Scott Wolk\altaffilmark{1},
 Andreas Zezas\altaffilmark{1}
}

\altaffiltext{1}{
 Harvard-Smithsonian Center for Astrophysics, Cambridge, MA 02138; mhicken@cfa.harvard.edu}
\altaffiltext{2}{
 Space Telescope Science Institute, Baltimore, MD 21218}
\altaffiltext{3}{
 NIST:  National Institute of Standards and Technology, Gaithersburg, MD, 20899}
\altaffiltext{4}{
 Department of Physics and Astronomy, University of Pittsburgh, Pittsburgh, PA 15260}
\altaffiltext{5}{
 Department of Astrophysical Sciences, Princeton University, Princeton, NJ, 08542}
\altaffiltext{6}{
 NASA:  Goddard Space Flight Center, Greenbelt, MD 20771}
\altaffiltext{7}{
 Department of Physics, Massachusetts Institute of Technology, Cambridge, MA, 02139}
\altaffiltext{8}{
 Planetary Science Institute, 1700 E. Fort Lowell Rd., Tucson, AZ 85719}
\altaffiltext{9}{
 Department of Physics, University of South Dakota, Vermillion, SD 57069}
\altaffiltext{10}{
 Department of Physics, Harvard University, Cambridge, Ma 02138}

\begin{abstract}
We present multi-band optical photometry of 94 spectroscopically-confirmed
Type Ia supernovae (SN Ia) in the redshift range 0.0055 to 0.073,
obtained between 2006 and 2011.  There are a total of 5522 light curve points.
We show that our natural system SN photometry has a precision of
$\lesssim0.03$ mag in \emph{BVr'i'}, $\lesssim0.06$ mag in $u'$, and
$\lesssim0.07$ mag in $U$ for points brighter than 17.5 mag and estimate 
that it has a systematic uncertainty of 0.014, 0.010, 0.012, 0.014, 0.046, 
and 0.073 mag in \emph{BVr'i'u'U}, respectively.  Comparisons of our 
standard system photometry with published SN Ia light curves
and comparison stars reveal mean agreement across samples in the range of
$\sim$0.00-0.03 mag.  We discuss the recent measurements of our 
telescope-plus-detector throughput by direct monochromatic illumination 
by \citet{cramer12}.  This technique measures the whole optical path through 
the telescope, auxiliary optics, filters, and detector under the same 
conditions used to make SN measurements.  Extremely well-characterized 
natural-system passbands (both in wavelength and over time) are crucial for 
the next generation of SN Ia photometry to reach the 0.01 mag accuracy level.
The current sample of low-z SN Ia is now sufficiently large to remove most of
the statistical sampling error from the dark energy error budget.  But pursuing
the dark-energy systematic errors by determining highly-accurate detector 
passbands, combining optical and near-infrared (NIR) photometry and spectra,
using the nearby sample to illuminate the population properties of SN Ia,
and measuring the local departures from the Hubble flow will benefit from 
larger, carefully measured nearby samples.
\end{abstract}


\section{Introduction}

The idea of using supernovae as tools for measuring the properties of cosmic 
expansion has a long history \citep{kirshner10}.
Pioneering work by the Calan/Tololo survey produced the first large sample of 
SN Ia light curves, with 29 SN Ia light curves measured with CCD detectors 
\citep{hamuy96b}.  Insight from Mark Phillips helped sharpen the use of 
SN Ia for distance determinations \citep{phillips93, hamuy96a}.  At the 
Harvard-Smithsonian Center for Astrophysics (CfA), we have been engaged in building up the local sample, 
with 22 SN Ia light curves in the CfA1 sample \citep{riess99}, 44~in 
CfA2 \citep{jha06} and 185~in CfA3 \citep[][hereafter H09]{hicken09a}.  
Additionally, Krisciunas and colleagues have published a
significant number \citep{krisc00, krisc01, krisc03, krisc04a, krisc04b,
krisc06}, the European Supernova Collaboration has published photometry of
various nearby SN Ia \citep[see][and references therein]{stanishev07, 
eliasrosa08, taubenberger08}, and \citet{kowalski08} published 
eight nearby SN Ia.
More recently, the Sloan Digital Sky Survey (SDSS) published 19 SN Ia with $z<0.100$ \citep{holtzman08} as part of the SDSS-II program
and the Lick Observatory Supernova Search (LOSS) published optical photometry
of 165 SN Ia \citep{mohan}.  The Carnegie Supernova Project (CSP)
produced 35 SN Ia light curves in its first release
\citep[][hereafter CSP1]{contreras10} and 50~in its second
\citep[][hereafter CSP2]{stritzinger11}, with a majority of
these objects including NIR photometry.

Looking to the future, CfA, LOSS and CSP continue building further nearby
samples.  Of particular interest will be $\sim$100 CfA NIR light curves 
that have optical photometry from CfA3 and CfA4 \citep{woodvasey08, 
friedman12}. 
In conjunction with this NIR and optical photometry, the CfA Supernova 
Group has taken spectra of many 
of these SN \citep{matheson08, blondin11, blondin12} 
using the FAST spectrograph \citep{fabricant98}.  

From its second and third years, SDSS-II should have photometry of 
another few dozen spectroscopically-confirmed SN Ia at $z<0.1$. 
The Palomar Transient Factory has discovered and spectroscopically confirmed
$\sim$900 SN Ia in its first two 
years\footnote{http://www.astro.caltech.edu/ptf/} and a large 
number of these should have 
quality light curves.  The Nearby Supernova Factory \citep{aldering02} has made 
spectrophotometric observations of several hundred SN Ia.  
ESSENCE will shortly publish their whole data set and cosmological analysis
\citep{narayan12}.  
PanSTARRS\footnote{http://pan-starrs.ifa.hawaii.edu/public/science-goals/active-universe.html} 
is also producing large numbers of SN Ia light curves that span the range
from low-z to cosmologically telling redshifts with a single photometric
system that should diminish the photometric uncertainties encountered by
splicing together separate samples at low redshift and high.

In addition to measuring SN data, we have also been engaged in improving 
the methods for determining SN Ia 
distances, using observations in multiple photometric bands to estimate both 
the luminosity of a SN Ia and its extinction \citep{riess96, jha07}.  MLCS2k2
is the most recent incarnation.  SALT2 \citep{guy07} and SiFTO \citep{conley08}
are other popular light curve fitters.
The current state of the art uses a 
statistical model for light curve shapes that predicts distances to $7\%$ for 
well-observed SN Ia using optical data and $5\%$ when NIR photometry 
is added \citep{mandel11}.

The application of the published nearby samples to cosmology includes the determination of 
$H_o$ to the $3\%$ level from the intercept of the Hubble diagram 
\citep{riess11}, where over half of the SN Ia sample came from CfA3.  
In addition, the photometry for two of the eight Cepheid-calibrated SN Ia 
(SN 2007af and SN 2007sr) came from the CfA3 sample and the improved 
photometric calibration of the comparison stars of a third
(SN 1995al) was obtained from observations taken at the 
F. L. Whipple Observatory (FLWO) 1.2m telescope 
during the course of the CfA3 observations \citep{riess09}.
Measurement of $q_o$ has been a major application that requires 
both high and low redshift samples that are on a common photometric system and
analyzed in a consistent way.  The pioneering work of measuring distant
SN Ia, first published by \citet{riess98} and 
subsequently by \citet{perlmutter99}, led to the surprising result of 
cosmic acceleration: $q_o < 0$.  Until 2009, the paucity of low-redshift SN Ia 
was a significant contributor to the statistical uncertainty in dark energy 
properties.  This changed in 2009, when \citet{hicken09b} used the CfA3 
data and data from the literature to construct the Constitution sample of 
SN Ia redshifts and distances, and employed it to improve constraints on 
the dark energy equation of state parameter, $w$.  The CfA3 sample was also 
instrumental in providing the data for \citet{kelly10} who first detected 
the small but real relation between SN Ia host-galaxy masses and the residuals 
from the distance predictions based on SN Ia light curves.  This was confirmed 
at higher redshift by work based on the Sloan SN survey \citep{lampeitl10} 
and on the Supernova Legacy Survey \citep{sullivan10}.  
A constraint on $\sigma_8$, the amplitude of cosmic fluctuations, based in large 
part on the CfA3 sample, has recently been derived by \citet{turnbull12}, 
who used the velocity residuals in the nearby Hubble flow to determine 
the variance in the dark matter density on a spatial scale of 8 Mpc,
$\Omega^{0.55}_m \sigma_{8,lin}=0.40\pm0.07$.

\citet{amanullah10} presented the Union 2.0 set of SN Ia distances, 
incorporating the CfA3 and first-year SDSS-II phototomery \citep{holtzman08}
while \citet{suzuki11} added 16 cosmologically-useful SN Ia from the
Hubble Space Telescope Cluster Supernova Survey, with 10 at $z>1$, to form 
the Union 2.1 set 
and provide tight limits on constant $w$ in a flat CDM model:  
$1+w = -0.013^{+0.068}_{-0.073}$, where the uncertainties include
all statistical and systematic errors.  \citet{sullivan11} combine all three
years of the Supernova Legacy Survey data with the nearby, first-year
SDSS-II and Higher-z \citep{riess07} samples to measure 
$1+w = 0.061^{+0.069}_{-0.068}$, where a flat universe is assumed and the 
uncertainties include all statistical and SN Ia systematic errors.
The Union 2.1 and Supernova Legacy Survey measurements of $w$ are the 
state of the art at the moment.

Since systematic errors in dark energy properties are now equal to or larger 
than the errors produced by finite sample sizes, progress demands ways to 
decrease those systematic errors.  Two of the largest sources of systematic 
uncertainty in using SN Ia for
cosmology are 1) the entanglement of intrinsic SN Ia color and host galaxy
reddening and 2) the overall accuracy of SN Ia photometry, especially in
accurately characterizing the passbands used.  

One promising path to 
overcoming the first source of systematic error is through NIR 
observations of SN Ia.  As shown by \citet{krisc04c}, \citet{woodvasey08}, 
CSP1, CSP2, and \citet{mandel11}, SN Ia are better standard candles in the 
NIR and 
extinction by dust is a less vexing problem.  The combination of the
optical and NIR photometry from the CSP and the CfA will lead to improved
disentangling of SN Ia color and host reddening.  Another interesting path to 
better distances and better understanding of the nature of SN Ia comes 
through combining information from light curves with information from 
spectra.  Recent work by \citet{blondin11} shows that spectra can be used 
to determine the intrinsic luminosity of SN Ia from the CfA samples.  This
builds upon findings by \citet{nugent95}, \citet{bongard06}, 
\citet{hachinger08} and 
\citet{bailey09}.  \citet{silverman12} combine nearby SN Ia spectra and
photometry to achieve the largest reduction in the Hubble diagram scatter 
via adding spectra to date.
Unlike the broad cosmological problem, the newer areas of combining
NIR photometry and optical spectra with optical photometry have 
significantly less data. 
This paper, with a sample of 94 new optical light curves, has substantial 
overlap with CfA NIR photometry \citep{friedman12} and optical spectra 
\citep{blondin12} 
of the same objects.  We expect that these measurements will be 
valuable in developing the tools to improve our knowledge of SN Ia and 
of the expansion history of the universe.

The second large source of systematic uncertainty will be greatly reduced 
through better characterization of detector passbands, as was done for our 
measurements by \citet{cramer12} using a monochromatic source to determine 
the system throughput, as described below.
In addition to better passband characterization, the significant overlap 
of nearby SN Ia photometry in the published CfA, LOSS, CSP and
other samples will aid in better understanding any possible 
systematic photometric effects that a given sample might have.

In \S2, we describe our observational and data reduction procedures.  
Greater emphasis is placed on the few differences as compared with
the CfA3 procedures and
a briefer description is provided where the procedures remained the same.
We then present the CfA4 light curves.
In \S3, we compare the overlapping objects betweeen CfA4, LOSS and CSP.
The CfA4 light curves, comparison star magnitudes and passbands can be found at
our website\footnote{http://www.cfa.harvard.edu/supernova/CfA4} or in the
online journal version of this paper. 
 
\section{Data and Reduction}

The CfA4 sample consists of 5522 light curve points.  All 94 SN have $BVr'i'$
measurements while 14 have $U$ and 12 have $u'$.  The average number of light
curve points per SN is 14.9~in $Vr'i'$, 12.5~in $B$, 7.8~in $u'$ and 7.0~in 
$U$.  The closest redshift, $z_{\rm CMB}$, is 0.0055 and the farthest is 0.073.  
The median redshift is 0.029 while the $25^{th}$ and $75^{th}$ 
percentile redshifts are 0.017 and 0.038, respectively.  89 of 94 SN 
have $z_{\rm CMB} > 0.010$.

CfA4 data processing followed the same three steps used for CfA3: 
reduction, calibration and host-galaxy subtraction.  Here we provide a 
brief overview of the overall process (see H09 for a more detailed treatment)  
and describe the differences between CfA4 and CfA3 in greater depth.
The reduction and subtraction
stages were carried out by a version of the ESSENCE and SuperMACHO pipeline
\citep{miknaitis07, rest05, garg07}.  
Calibration paralleled CfA3 but was more automated.  We employed 
differential photometry, 
calibrating the comparison stars surrounding the SN on photometric nights 
and then measuring the flux of the SN relative to the comparison stars in 
each data image, on both photometric and non-photometric nights.  
We employed host-galaxy subtraction for all 94 SN, using multiple
reference images for the majority of the SN.

\subsection{Instruments}

The CfA4 data were obtained on the 1.2m telescope at the 
FLWO using the single-chip, four-amplifier CCD
KeplerCam\footnote{http://linmax.sao.arizona.edu/FLWO/48/kepccd.html}.
Observations were acquired on amplifier two with a pixel scale of $0".672$,  
resulting in a field of view of approximately 11'.5x11'.5.  
The 1.2m primary mirror deteriorated during the course of
the CfA4 observing and its effects will be described below.  A replacement
mirror is nearly ready.

Due to the KeplerCam's good cosmetics, a bad-pixel mask was not required.
The same \emph{BVr'i'} filters from CfA3 were used for CfA4.  The second 
of the two CfA3 $U$ filters was used for CfA4
until it broke in January, 2009 and afterwards an SDSS $u'$ filter was used.  
A further description of the filters used in conjunction with the KeplerCam
can be found at the FLWO 
website\footnote{http://linmax.sao.arizona.edu/FLWO/48/CCD.filters.html}.

\subsection{Observations}

The CfA Supernova Group depends on both professional and amateur SN
searches for its observing targets.  
Most of these search surveys had typical limiting magnitudes of 19.5 mag. 
The 1.2m telescope can reach targets north of declination $-20^\circ$.  The
CfA4 discovery data are displayed in
Table \ref{table_sndiscovery}.  The reported SN positions are a weighted mean
from our subtracted images, usually in $r'$, and are usually an improvement
over the announced discovery positions.  These positions will be of use for 
studies requiring more accurate SN positions, such as exploring the host galaxy 
properties at the point of explosion.  For the reader's convenience we also 
list the redshift, host galaxy name, and Milky Way color excess for each SN.

As explained in H09, the CfA Supernova Group rapidly acquires spectra of many
of the new SN brighter than $\sim18.5$~mag and northwards of $-20^\circ$ 
to provide typing and follow-up investigation.  We also begin taking optical 
and $JHK$ photometry.  This combination allows for a richer understanding of
both individual SN and the sample as a whole.  Priority was usually given to 
younger and more interesting SN.

The SN in our sample come from a variety of SN searches.  In 
many cases, the SN are detected in galaxies which are targeted for 
monitoring.  This means that the host galaxies do not constitute an unbiased 
sample of the universe, and the properties of the SN in this sample 
and of their hosts are not necessarily representative.  See H09 
for further details.


\subsection{Pipeline:  Reduction Stage}

Images underwent bias subtraction and flat fielding.  Dome-screen flats were 
used for \emph{BVr'i'} while twilight flats were used for $Uu'$.  The $i'$-band 
fringes were slightly larger than in CfA3 so fringe corrections were applied.
Cosmic rays were removed in the same way as in CfA3.

The UCAC3 catalog \citep{ucac3} was used to produce a linear astrometric 
solution for the vast majority of the CfA4 images.  The USNO-B1.0 
\citep{monet03} or USNO-A2.0 catalogs \citep{monet98} were employed in the
few cases where the UCAC3 catalog was too sparse.  SWarp
\citep{bertin02} was run to properly scale and align the images.  
DoPHOT \citep{schechter93} was then used to calculate fluxes for all 
stellar-shaped objects.

\subsection{Calibration}

We used \citet{landolt92} to calibrate our \emph{UBV} bands and 
\citet{smith02} to calibrate our \emph{r'i'} bands.  For our $u'$ calibration
we transformed the \citet{landolt92} $U$ magnitudes into $u'$ via the
equation $u'=U+0.854$ \citep{chonis08}.

As in CfA3, we performed aperture photometry on the Landolt/Smith standard 
stars and on our SN-field comparison stars using the NOAO/DIGIPHOT/APPHOT 
package in IRAF \citep{tody93}.  Comparison stars were chosen so that 
they were reasonably well isolated.  Due to the deteriorated mirror, which
resulted in larger stellar point spread functions (PSF), an aperture with
radius of 18 pixels was used on both the standard and comparison stars.  This
was larger than the 15 pixels used in CfA3.   An
aperture correction was calculated from as many as four bright and isolated
stars by subtracting the 6-pixel-radius-aperture magnitude from
the 18-pixel-radius-aperture magnitude and applied to the 6-pixel-radius
magnitude of all of the stars in the field.

As in CfA3, a linear photometric transformation solution for each night 
was calculated from our Landolt/Smith stars using system of equations 1.  

\begin{eqnarray}
u-b &=&\rm zp_{UB} + \alpha_{UB}x + \beta_{UB}(U-B) \nonumber \\
b-v &=&\rm zp_{BV} + \alpha_{BV}x + \beta_{BV}(B-V)  \nonumber\\
v-V &=&\rm zp_{V} + \alpha_{V}x + \beta_{V}(B-V)  \nonumber\\
v-r &=&\rm zp_{Vr'} + \alpha_{Vr'}x + \beta_{Vr'}(V-r')  \nonumber\\
v-i &=&\rm zp_{Vi'} + \alpha_{Vi'}x + \beta_{Vi'}(V-i')
\end{eqnarray}

The terms on the left side of the equations are the instrumental colors except
for the \emph{V}-band term.  The first term on the right side of each equation
is the zero-point, followed by the airmass coefficients, $\alpha$, times the
airmass, $x$.  The \emph{V}-band equation is unique in that it directly
relates the instrumental magnitude \emph{v} to the standard system magnitude
and color, \emph{V} and $B-V$.  The other four equations only relate the
instrumental and standard-system colors to each other.  The final term on the
right of the four color equations multiplies the standard-system color of the
standard stars by a coefficient, $\beta$, to convert the standard-system color
into the natural-system color.  When the $u'$ filter replaced
the $U$ filter we used the above equations replacing $U$ with $u'$.

The photometric solution was applied to the comparison star measurements.  
This produced tertiary standards that were used to
calibrate the SN measurements in the natural system.  
To calculate the photometric zero-point
for each SN image, we took a weighted mean of the differences between our
calibrated magnitudes and the instrumental DoPHOT measurements of the
comparison stars.  

Most of our SN fields were observed on multiple photometric nights to ensure
more accurate calibration.  However, this was not always possible.  All 
SN fields that were calibrated on only one night had other SN fields calibrated
on the same night that were consistent across multiple nights.  This 
increases confidence but does not guarantee the single-night calibrations.
The comparison star uncertainties include the measurement uncertainties, the 
standard deviation
of measurements from multiple nights (for single nights, an appropriate error
floor was used instead) and the uncertainty of the transformation to the
standard system.  The typical uncertainty of our \emph{V}-band comparison star
measurements is 0.015 mag.  

\begin{figure}
\scalebox{0.80}[0.80]{
\plotone{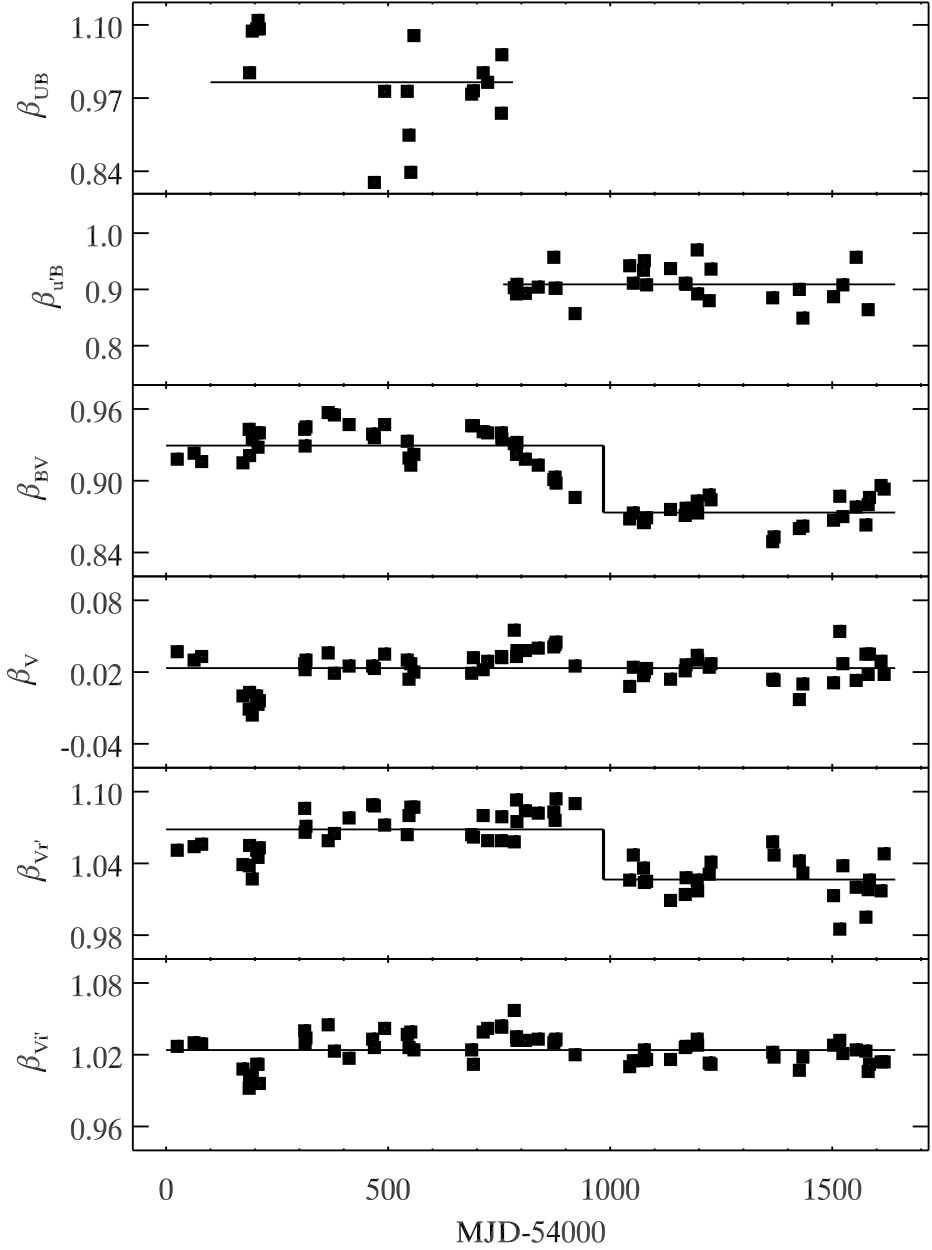}
}
\caption{$Uu'BVr'i'$ color coefficients are plotted versus time with
the average value over the relevant time periods shown as a solid line.  The
$v-V$ and $V-i'$ coefficients are sufficiently stable to be represented
by one constant value across the whole time domain while the $B-V$ and
$V-r'$ coefficients are each better described by one value in period one
and another value in period two.  The $U-B$ coefficients have large
uncertainties (not shown) and a large scatter while the $u'-B$ coefficients
have a much smaller scatter, illustrating the superior precision of $u'$
measurements.
}
\label{fig_colterms}
\end{figure}

The color coefficients from each photometric night are plotted in Figure 
\ref{fig_colterms}.  The $V$, 
$V-i'$, $U-B$ and $u'-B$ coefficents do not show any significant trend over
time and can be fit well by one average value while the $B-V$ and $V-r'$ 
coefficients show a step-function distribution with one value in
the period before mid-2009 (period one) and another value in the 
period after (period two).  We chose to use 2009 August 15 (MJD=55058) as the
dividing point and calculated average $B-V$ and $V-r'$ color coefficients
for period one and another average for period two.  The color coefficients 
are listed in Table \ref{table_colorterms}.  The largest difference between the
two periods was in the $B-V$ color coefficients, which decreased from $0.93$ 
to $0.87$.  Since the $V$-band color coefficient was stable across the two 
time periods this implies that the $B$ passband shifted redward in period 
two.  Deposits or condensation on the camera are likely causes for 
the changing color coefficients.  The KeplerCam was baked
on 2011 May 17 to remove deposits/condensation and the $B-V$ and $V-r'$ 
color coefficients derived after this returned to their period-one values.  
However, these 
post-baking photometric nights were not used for any CfA4 photometric 
calibration so only the two time periods, one before and one after August, 
2009, with their respective color coefficients were needed.  
In addition to this, the 1.2m mirror was deteriorating from 2007
to 2011, losing about 0.6 mag of sensitivity in $V$.  We note that the 
KeplerCam underwent regular bakeouts every August, but none of these produced a 
dramatic difference like the one in 2011 May.  There is no evidence that the 
2011 bakeout procedure was different from previous cycles of dessication and 
cleaning, but the result was a significant change in the color coefficients.  
Whatever the cause, the $B-V$ and
$V-r'$ color coefficients returned to their period-one values and the most
likely explanation is that the 2011 May bakeout removed deposits/condensation
that previous ones could not.  

Synthesized natural system \emph{BVr'i'} passbands for the KeplerCam were
calculated in H09 by
combining the primary and secondary mirror reflectivities (taken as the square
of the measured reflectivity of the primary), the measured filter
transmissions, and the measured KeplerCam quantum efficiencies.  No atmospheric
component was included.  These passbands were presented as normalized photon
sensitivities.  No $U$-band passband was made due to a lack of a $U$
filter transmission curve.

More recently, \citet[][hereafter C12]{cramer12} measured the FLWO 1.2m 
KeplerCam 
$BVr'i'$ passbands using the technique initially described in
\citet{woodward10} and \citet{stubbs06}.  The dome screen was 
illuminated with light from a
monochromatic, tunable source to generate a series of monochromatic dome flats
spanning each filter passband.  A NIST-calibrated 
photodiode\footnote{http://www.nist.gov/calibrations/upload/sp250-41a.pdf} 
monitored the total amount of light incident on the telescope during each 
exposure.  Filter passbands were generated by scaling the camera response 
to the photodiode signal.  The measured passbands therefore include not 
only the filters, camera, and mirrors, but also the effect of other optics 
in the telescope optical train -- notably, a doublet corrector lens with 
an aging anti-reflective coating on each of the four surfaces -- as
well as accumulated dirt and condensation in the camera.  The Vr'i' 
passbands were measured four times: in July and October of 2010, and 
April and June of 2011.  The
B passband was measured in 2010 October as well as April and June of 2011.  
We therefore have measured passbands both before and after the 2011 May 
bakeout.  For more details and tables of the measured passbands, see C12.

\begin{figure}
\epsscale{0.7}
\plotone{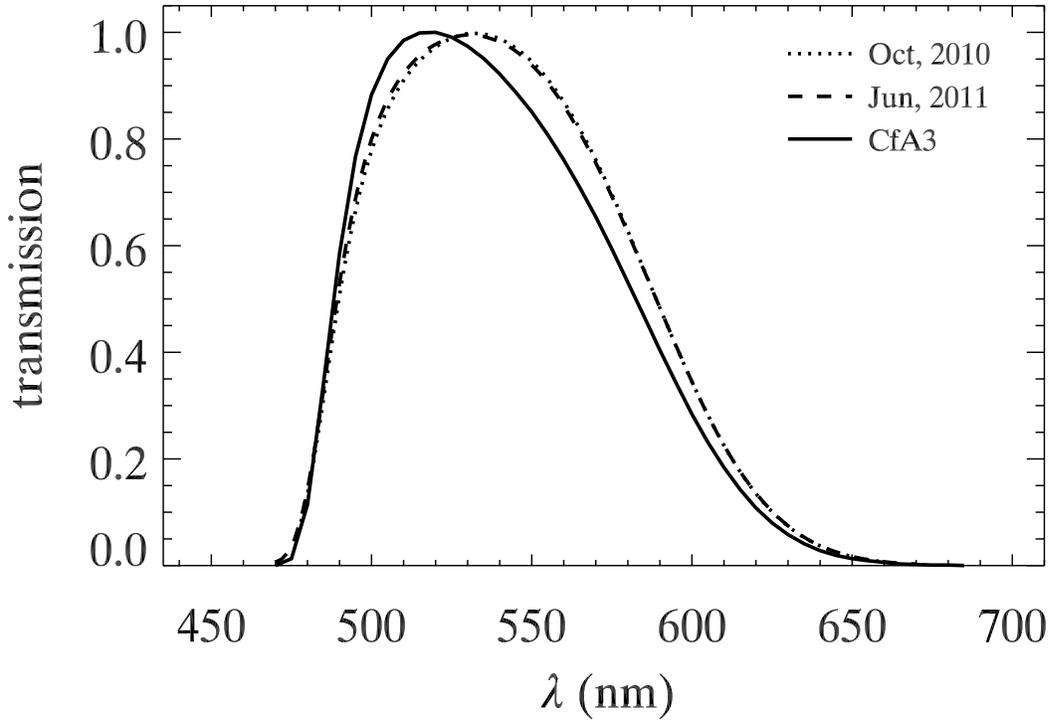}
\caption{The C12 $V$ passbands measured before (dotted) and after (dashed) 
the 2011 May bakeout are highly consistent with each other and in 
reasonable agreement with the CfA3 synthetic $V$ passband (solid).
}
\label{fig_V_passband}
\end{figure}

\begin{figure}
\epsscale{0.7}
\plotone{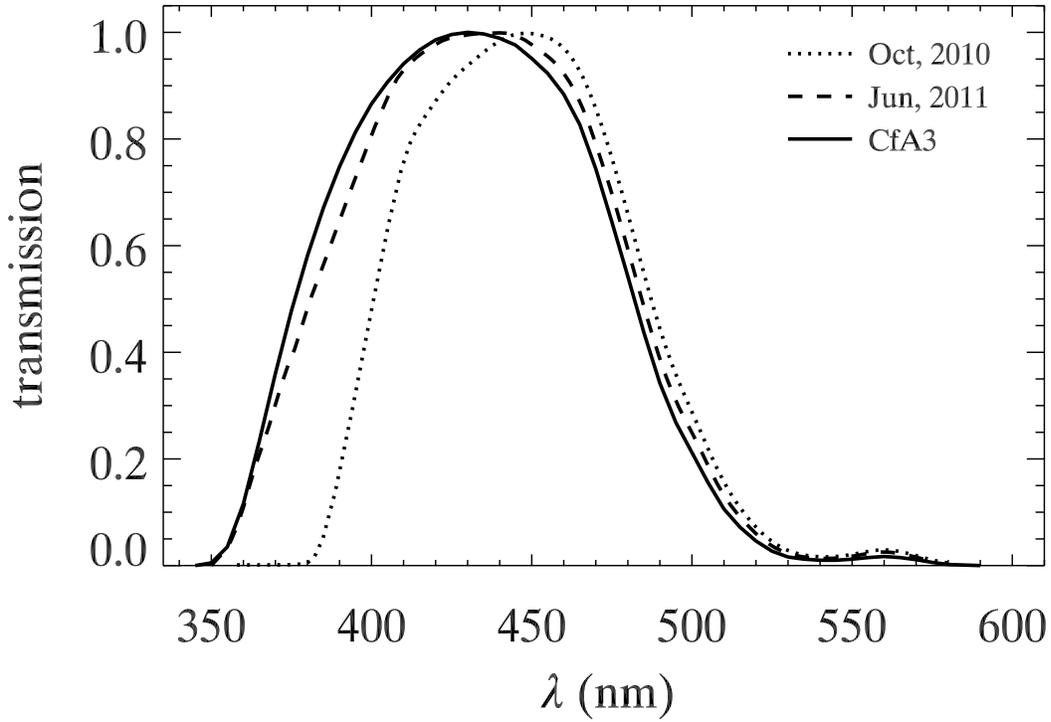}
\caption{The C12 $B$ passbands measured before (dotted) and after (dashed) 
the 2011 May bakeout, showing the blueward shift caused by 
the bakeout.  The CfA3 synthetic $B$ passband (solid) agrees fairly well
with the post-bakeout C12 $B$ passband.  
}
\label{fig_B_passband}
\end{figure}

The C12 $Vr'i'$ passbands are relatively 
stable across the bakeout and agree reasonably well with the synthesized CfA3 
$Vr'i'$ passbands (see Figure \ref{fig_V_passband} for $V$) 
but the C12 pre-bakeout $B$ passband is significantly redward
of both its post-bakeout counterpart and the synthesized CfA3 $B$ passband,
as seen in Figure \ref{fig_B_passband}.  The pre-bakeout C12 passbands 
were observed at two separate times, six months apart, and were virtually 
identical.  The period-two color coefficients were also stable over this 
same range of time (and longer).  The implication is that the pre-bakeout 
C12 passbands are valid over the period-two time range where the 
deposits/condensation were present.  

After the bakeout, the C12 $B$ passband shifted bluewards while the C12 
$V$ passband remained stable.  This is consistent with the increase of the 
$B-V$ color coefficients back to the period-one 
$0.93$ level.  The post-bakeout C12 
$B$ passband can also be seen to be much more consistent with the 
synthesized CfA3 $B$ passband.  
We also point out that the CfA3 $B-V$ color coefficient was $0.92$, fairly 
close to the CfA4 period-one value.  These pieces of information suggest 
that the post-bakeout C12 $B$ passband can be used as the CfA4 period-one 
natural system $B$ passband.

To summarize, the C12 post-bakeout passbands should be used in conjunction with
the CfA4 period-one natural-system light curves while the C12 pre-bakeout
passbands should be used for period two.  Also, due to the reasonable 
consistency of the $BVr'i'$ CfA3 and CfA4 period-one color coefficients, 
the C12 post-bakeout passbands can be used with the CfA3 natural-system 
light curves.  The similarity of the synthesized CfA3 passbands to the C12 
$Vr'i'$ and post-bakeout $B$ passbands suggests that they were sufficiently 
accurate and that their use in cosmological studies was satisfactory 
\citep{amanullah10, suzuki11, sullivan11}.

\subsection{Pipeline:  Host-Galaxy Subtraction}

Each of the 94 SN in the CfA4 sample underwent host-galaxy subtraction.  
Reference images were acquired on clear nights with good seeing and little 
or no moon so as to maximize their signal-to-noise ratio.  
However, due to the poor mirror 
quality the images had larger PSF sizes than in CfA3.  To combat this, multiple
reference images were subtracted from the majority of the data images.

The same subtraction algorithm and software as in CfA3 were used for CfA4. 
A convolution kernel that
transforms the PSF of one image to the PSF of the other was
calculated using the algorithm of \citet{alard98} and
\citet{alard00} with slight improvements as in \citet{becker04} and
\citet{miknaitis07} and then the subtraction was performed.  
The SN flux in the difference image was then measured with the DoPHOT PSF from 
the stars of the unconvolved image.  

The natural system flux normalization for the
difference image was chosen from the SN data image as opposed to the reference
image.  This ensured that the normalization of the SN data would
be consistent with the passbands and color coefficients from the same time
period.

Noise maps were propagated for both data and reference images and used 
to calculate a noise map for the difference image.  Information from the 
noise image was combined with the DoPHOT uncertainty and calibration 
uncertainty to produce the uncertainty of the natural system SN measurement.

The subtraction process was not always perfect and this introduced
extra uncertainty.  Steps were taken to estimate this uncertainty.
In the cases where multiple reference images existed and were successfully
subtracted from one data image we had a distribution of values that provided
a better estimate of the true SN flux and its 
uncertainty.  The differences in these
values were due to differences in the various reference images. 
In order to arrive at one final light-curve point, the multiple 
photometry values from each data image were plotted and any extreme 
outliers were removed.  Suppose that $N$ values remain.  They are 
different from each other due to Poisson noise in the host-galaxy and sky flux 
in the reference images and due to slight limitations in convolving
every reference image equally well to the data image.
We took the median of these $N$ values to be the final light-curve data point.  
There are also $N$ photometry-pipeline uncertainties associated with each
of the $N$ photometry values.
To calculate the final light-curve uncertainty for the data point in question,
$\sigma_{\rm total}$,  
the median of the $N$ photometry-pipeline uncertainties (we will call
this median $\sigma_{\rm pipe}$)
was added in quadrature to the standard deviation of the $N$
photometry values ($\sigma_{\rm phot}$).  

Formally, there is a slight double counting of the
Poisson noise of the reference images since it is part of the standard
deviation of the $N$ photometry values, $\sigma_{\rm phot}$,
 and is also included in the 
difference image noise maps.  However, this is dwarfed by the size of the
other uncertainties and has no significant effect on the size of the final 
error bars.  In $\sigma_{\rm phot}$ the Poisson noise 
of the reference image is typically much smaller than the uncertainties due 
to imperfect subtractions.  And in the pipeline uncertainty, the combination 
of the data image noise with the DoPhot and calibration uncertainties is 
larger than the reference image noise which is taken during dark time
with better seeing.  

For the cases where only one reference image was successfully subtracted the
light curve value was simply the single-subtraction value.  The uncertainty
from the single-subtraction photometry was added in quadrature to an estimate
of what the standard deviation would have been had multiple reference images
existed.  This estimate was based on a quadratic fit of the standard 
deviations of the
multiple-subtraction photometry values versus SN magnitude in a given band at
magnitudes fainter than 16 mag.  For magnitudes brighter than this a constant
but representative value was used.  This 
derived function then gave a reasonable estimate that was easily calculated
from the SN magnitude of the single-subtraction point.  
This function is presented in Table \ref{table_singlerefimage}.  

Having multiple reference images for most of the CfA4 sample is one of the main 
differences with CfA3.  In CfA3, there was only one reference images for most
of the SN and none for some SN with far-away hosts.  The uncertainties for the
single reference-image CfA3 photometry were what are here called 
$\sigma_{\rm pipe}$ for $N=1$ and are almost certainly an underestimate of the
true uncertainty.  Future use of the CfA3 sample would benefit by adding
an estimate of $\sigma_{\rm phot}$.  The function we derived for the 
CfA4 single reference-image photometry could serve as such an estimate and can
be added in quadrature to the quoted CfA3 uncertainties.  

It was found that the standard deviation of the 
$N$ light-curve values was typically on the order of their photometry-pipeline 
uncertainties.  Using multiple reference images increased the 
accuracy of both the CfA4 light curve values and their uncertainties.
We believe that the final error bars are our best estimate
of the true uncertainties.

\begin{figure}
\hspace{-1.2in}
\scalebox{1.70}[1.70]{
\plotone{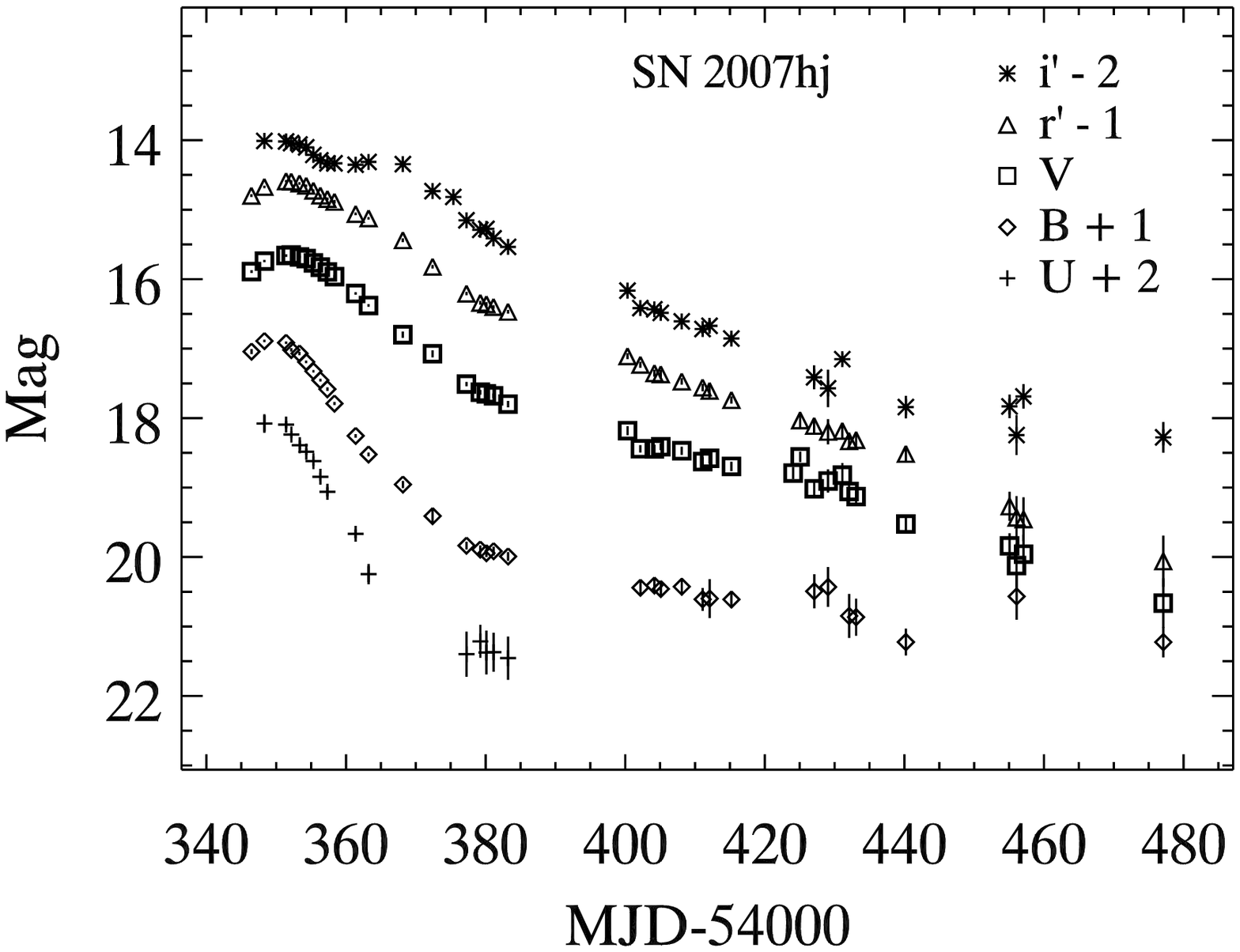}
}
\caption{SN 2007hj, one of the better-sampled CfA4 light curves.  The 
error bars are smaller than the symbols for most of the data points.}
\label{fig_lc1}
\end{figure}

\begin{figure}
\hspace{-1.2in}
\scalebox{1.70}[1.70]{
\plotone{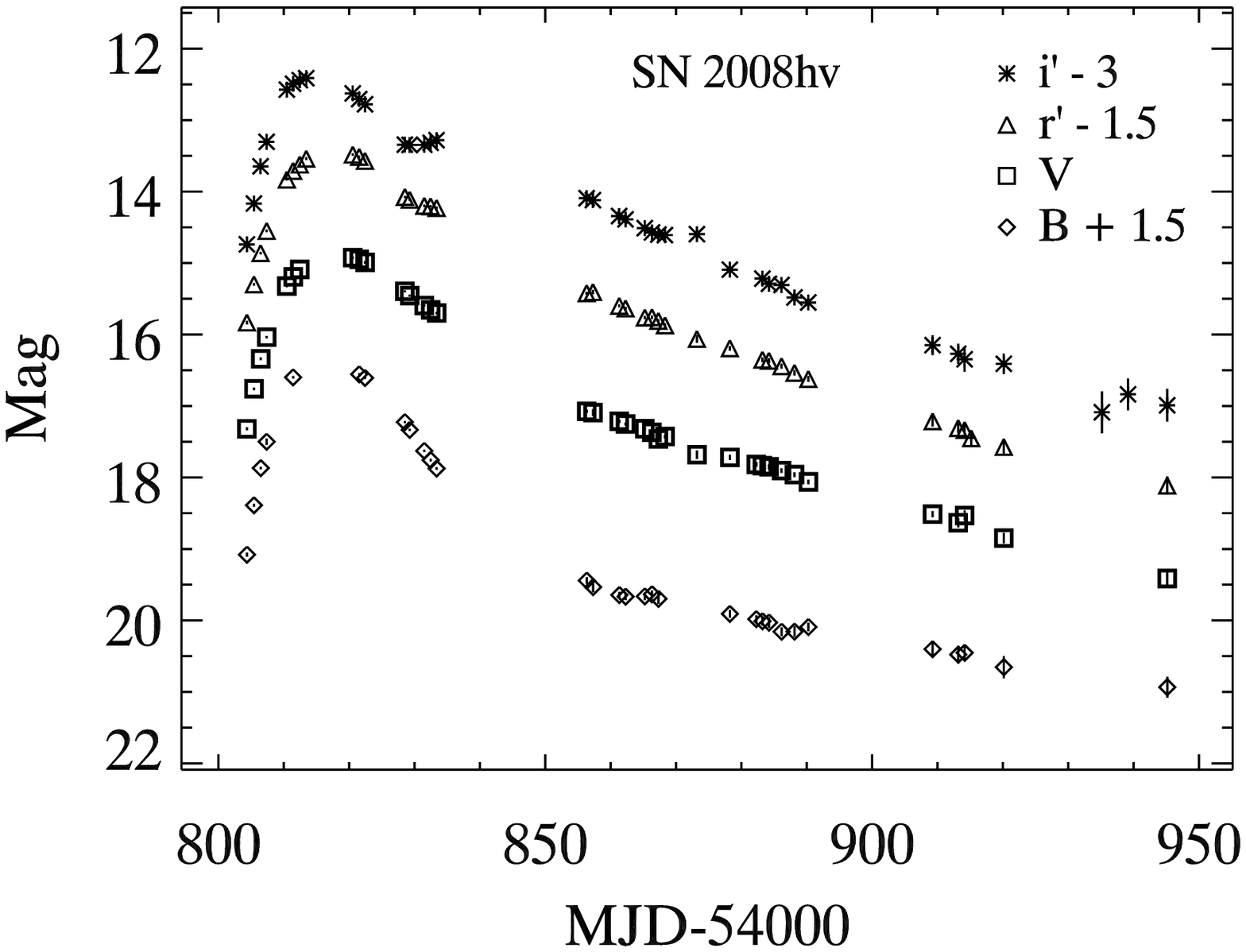}
}
\caption{SN 2008hv, one of the better-sampled CfA4 light curves.  The 
error bars are smaller than the symbols for most of the data points.}
\label{fig_lc2}
\end{figure}

\begin{figure}
\hspace{-1.2in}
\scalebox{1.70}[1.70]{
\plotone{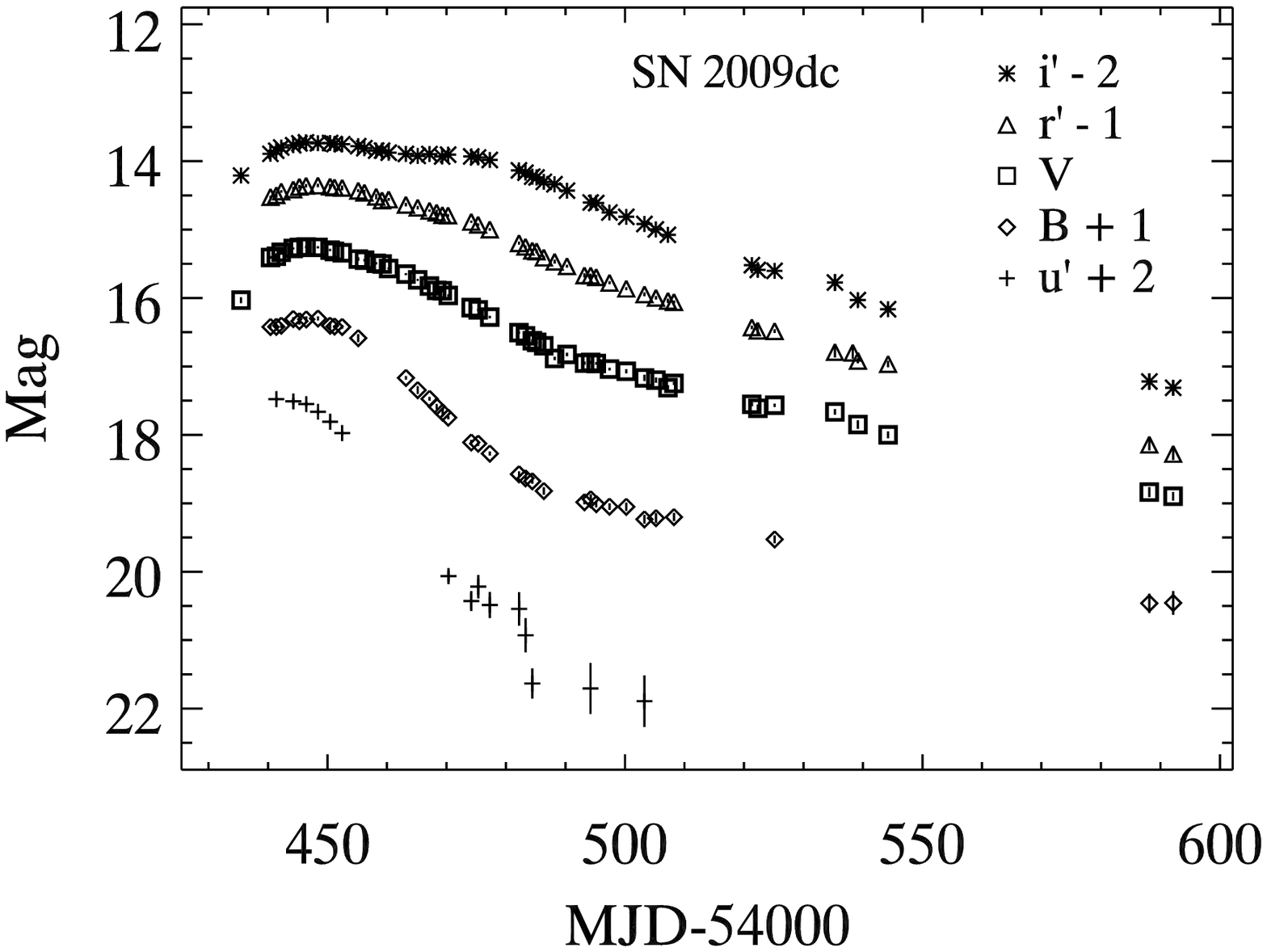}
}
\caption{SN 2009dc, one of the better-sampled CfA4 light curves.  The 
error bars are smaller than the symbols for most of the data points.}
\label{fig_lc3}
\end{figure}

\begin{figure}
\hspace{-1.2in}
\scalebox{1.70}[1.70]{
\plotone{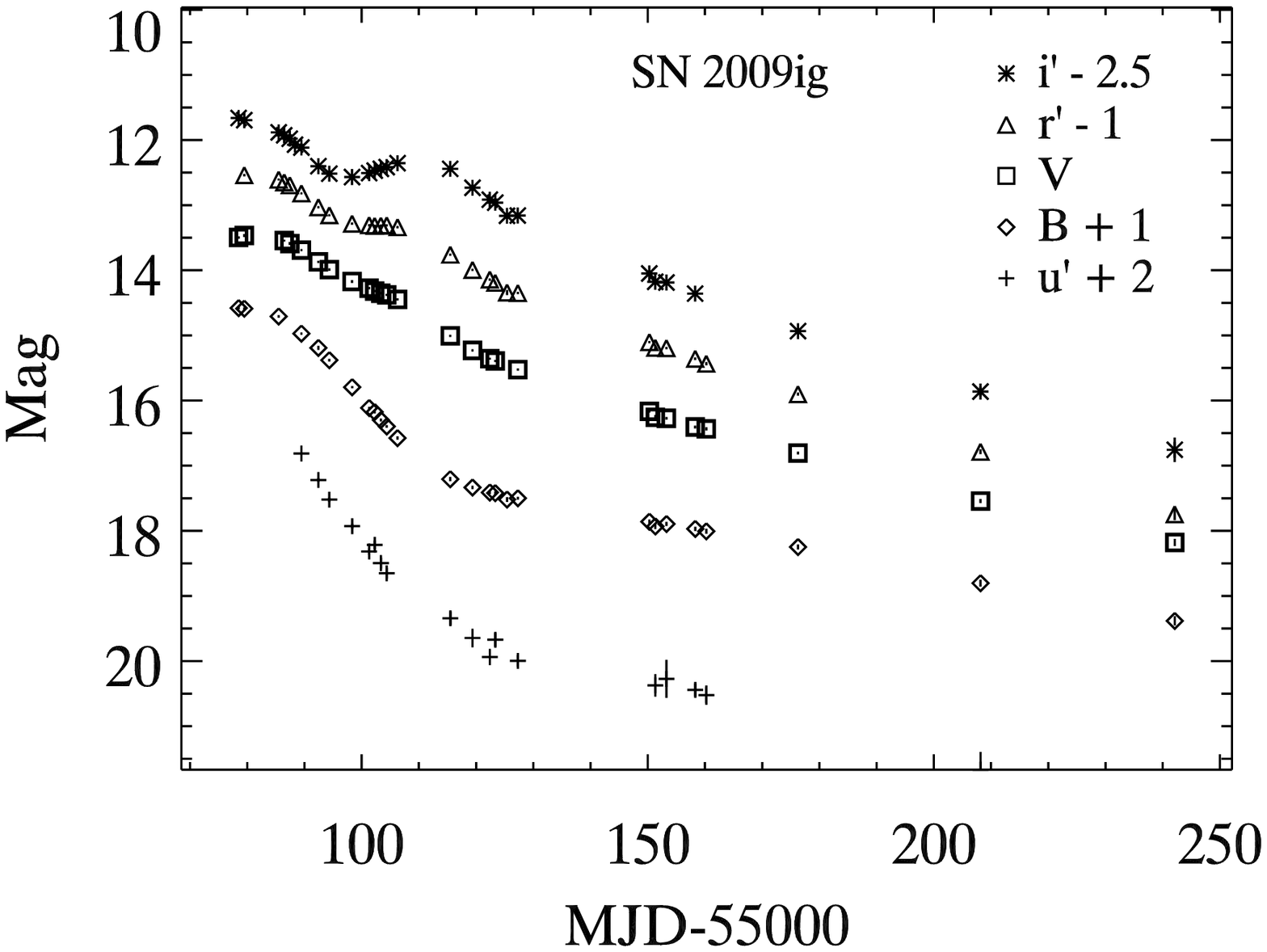}
}
\caption{SN 2009ig, one of the better-sampled CfA4 light curves.  The 
error bars are smaller than the symbols for most of the data points.}
\label{fig_lc4}
\end{figure}

Our light curves were produced in the natural system
and then converted to the standard system by using the color coefficients 
in Table \ref{table_colorterms}.  
Figures \ref{fig_lc1}, \ref{fig_lc2}, \ref{fig_lc3}, and \ref{fig_lc4} 
show four of the better-sampled CfA4 light curves.
The standard-system comparison stars and both natural
and standard system light curves for all the objects
can be found in the online version of Tables 
\ref{table_star_std}, \ref{table_lc_nat}, and \ref{table_lc_std}, and at our
website\footnote{http://www.cfa.harvard.edu/supernova/CfA4}. 
The print journal version and astro-ph version of Tables 
\ref{table_star_std}, \ref{table_lc_nat}, and \ref{table_lc_std} only
show a small portion of the full data set.  
The natural-system comparison star photometry
is also available at our website.  For the SN photometry, the number of
successful subtractions from a given night that survived outlier rejection 
is listed.  For example, if there were two data images and seven reference 
images and no photometry points were rejected then $N=14$.  Usually there was 
only one data image per night.  The uncertainties that were added in 
quadrature to obtain the final uncertainty are listed.  The C12 \emph{BVr'i'} 
passbands will be available soon.  The natural system passbands and 
photometry can be used together to avoid the uncertainty of using 
star-derived color coefficients for SN. 

We make no effort to estimate the
additional uncertainties in the standard-system SN Ia photometry due to the
lack of s-corrections but note that
the uncertainties listed in Table \ref{table_lc_std} 
are certainly an underestimate. 
The natural-system uncertainties are the same as the
standard-system uncertainties because we chose not to add the 
stastistical uncertainty of the color terms which would increase the total
uncertainty to about 1.005 times the natural-system values and are 
thus negligible.  As evident in Table \ref{table_lc_nat}, our natural system 
SN photometry has a precision, $\sigma_{\rm total}$, of
$\lesssim0.03$ mag in \emph{BVr'i'}, $\lesssim0.06$ mag in $u'$, and 
$\lesssim0.07$ mag in $U$ for points
brighter than 17.5 mag.  

We also estimate a systematic uncertainty 
in the natural system photometry of each SN of
0.014, 0.010, 0.012, 0.014, 0.046, and 0.073 mag in \emph{BVr'i'u'U}, 
respectively.  These systematic uncertainties
are not included in our natural system photometry uncertainties in Table
\ref{table_lc_nat}.  They are due to the 
uncertainties in the zero points of the photometric solution.  They were derived
by dividing the median uncertainty of all nights' photometric-solution 
zero points in a given band 
by the square root of the average number of nights of photometric calibration
across all SN in the same band.  For example, the median uncertainty in the
$V$ solution is 0.02 mag and the average number of nights is 3.6, resulting
in 0.01 mag.  These estimates are in rough agreement with the differences
between samples seen in Table \ref{table_compstars_sum}, which also serve as an
estimate of the systematic offsets as explained below.  
The comparison star uncertainties in Table 
\ref{table_star_std} contain this systematic uncertainty but the photometry 
pipeline treats them as purely statistical and so it gets lost due to the 
relatively large number of comparison stars used for each SN.    

\section{Photometry Comparison With Other Samples }

Twelve CfA4 SN Ia light curves overlap with recent LOSS photometry 
\citep{mohan, silverman11} and 
eight overlap with CSP2.  Comparisons between the three groups were made in
the standard system.  The LOSS and CSP2 comparison star photometry was
published in the standard system as was the LOSS SN photometry.  The CSP2 SN
photometry was only presented in the natural system.  The CSP2 SN standard 
system
photometry (without s-corrections) was provided to us for the overlapping 
objects (Stritzinger, M. 2011, private communication).  We emphasize that none
of the CfA4, LOSS or CSP2 standard-system SN photometry is s-corrected. 
Since SN spectral energy distributions (SEDs) differ from the stellar SEDs
used to derive the photometric transformation coefficients, the 
comparisons of the standard-system SN photometry here is limited to providing
a reasonable but not highly accurate idea of the agreement between samples.  
A description of s-corrections and why they are needed for more-accurate 
transformation to the standard system is given in 
\citet{suntzeff00}.  An application of this method is presented in 
\citet{stritzinger02}.

\begin{figure}
\includegraphics[scale=.80]{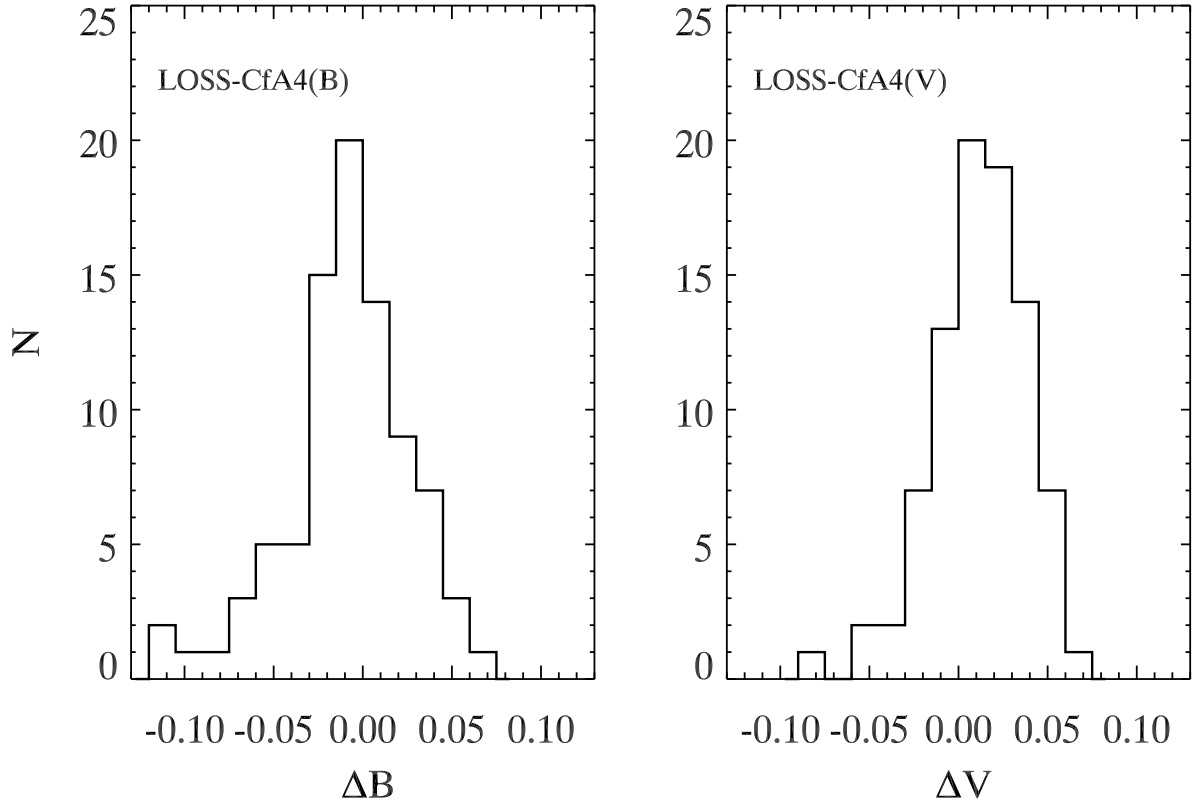}
\caption{Histograms of the differences in the LOSS-CfA4 $BV$ photometry 
for all the comparison stars in common between the two samples.  The 
distributions are roughly symmetric.}
\label{fig_loss-cfa4_star_diff}
\end{figure}

\begin{figure}
\vspace{3.0in}
\scalebox{1.10}[1.10]{
\plotone{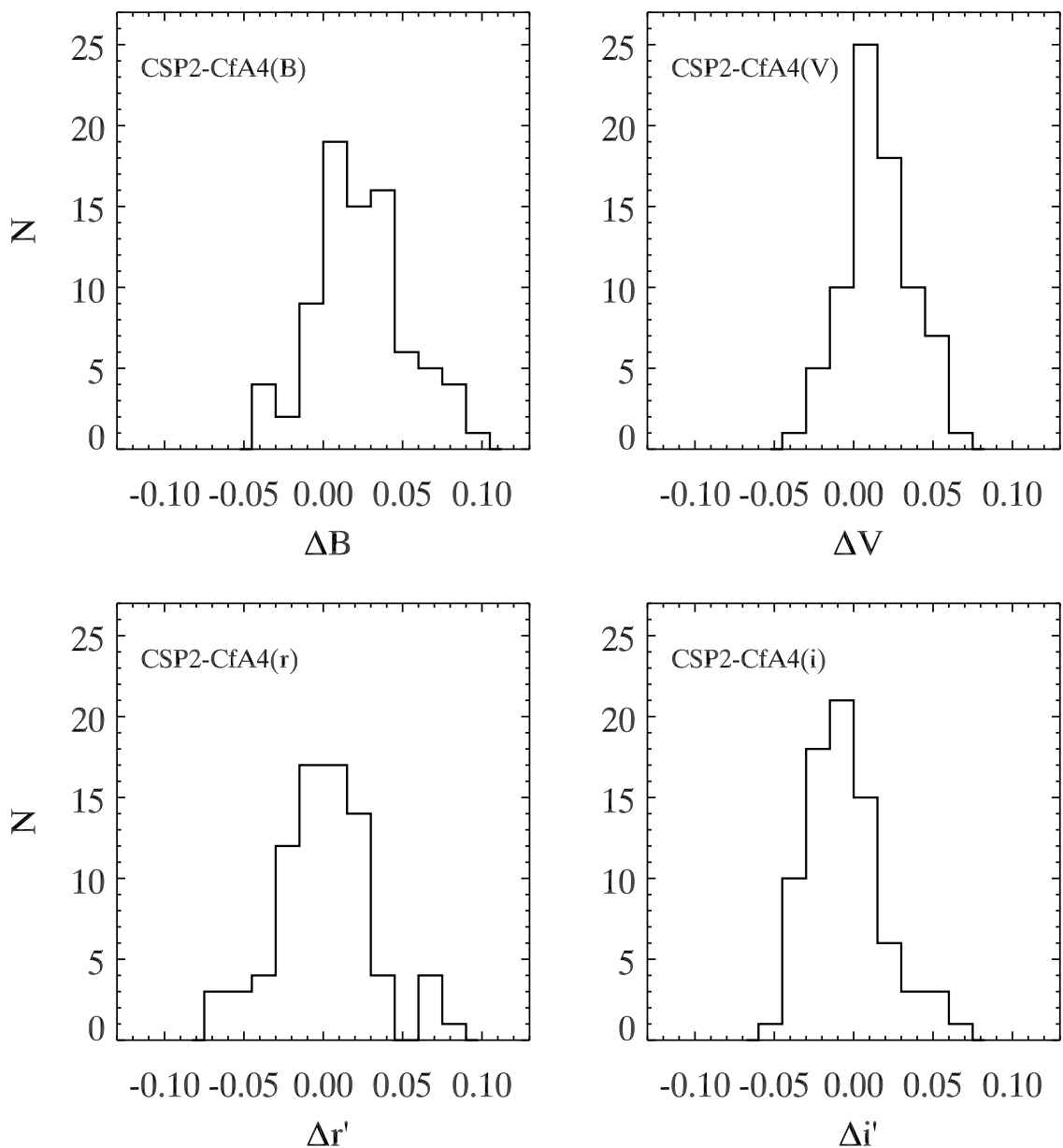}
}
\caption{Histograms of the differences in the CSP2-CfA4 $BVr'i'$ photometry 
for all the comparison stars in common between the two samples.  The 
distributions are roughly symmetric in $Vr'$ with slightly larger
tails on the positive side in $Bi'$.}
\label{fig_csp2-cfa4_star_diff}
\end{figure}

\begin{figure}
\includegraphics[scale=.80]{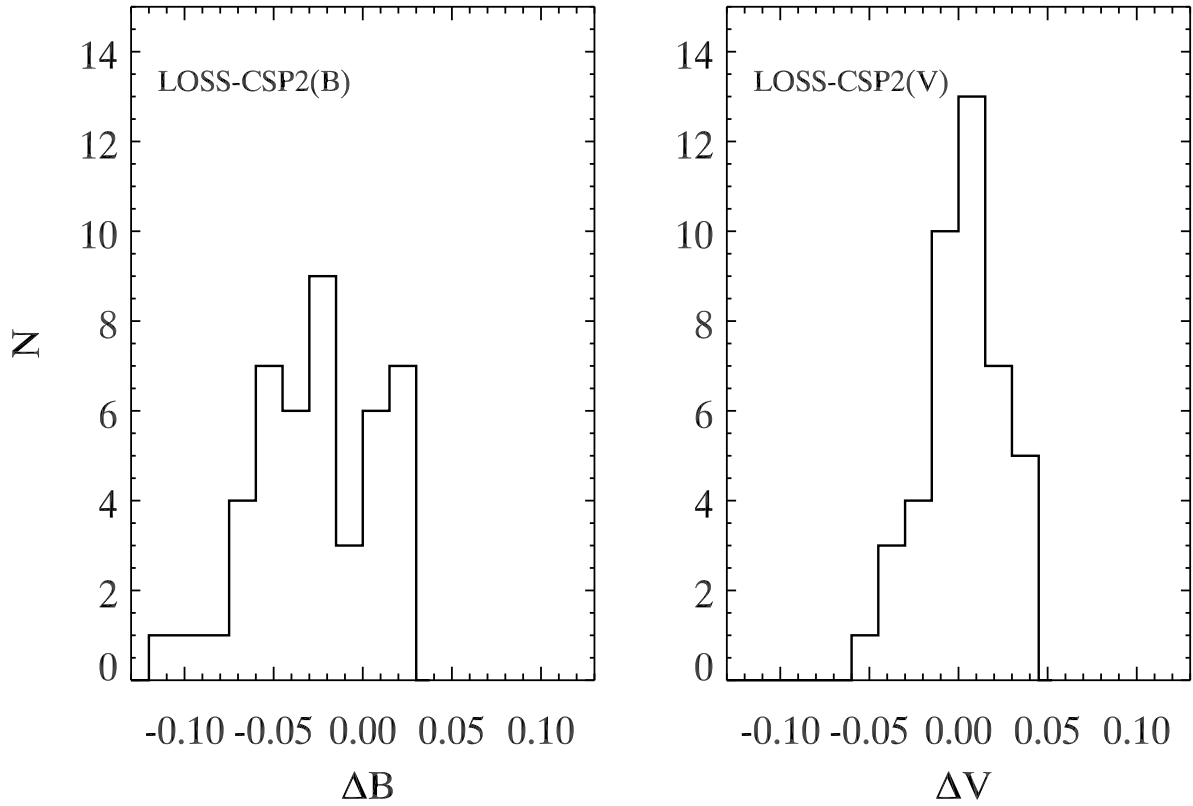}
\caption{Histograms of the differences in the LOSS-CSP2 $BV$ photometry 
for all the comparison stars in common between the two samples.  The 
distributions are roughly symmetric in $V$ but not in $B$.}
\label{fig_loss-csp2_star_diff}
\end{figure}

Table \ref{table_compstars_sum} shows the mean difference
of all the comparison stars in common between CfA4, LOSS and CSP2.  
There is relatively good agreement ($<0.015$ mag) in all bands and between
all samples
except in $B$ where CSP2 and LOSS differ by 0.035 mag and CSP2 and CfA4
differ by 0.022 mag.  The differences in the mean are larger than the 
standard error of the mean in all cases except for LOSS-CSP2 in $V$.  
The comparison star photometry does not require s-corrections and gives a good
idea of the systematic offsets that would exist between the three groups'
SN photometry after accurate s-corrections.  The mean difference between
two groups' photometry can be taken by itself as a good estimate of the 
systematic offset.  Another approach, in the spirit of having $\chi^2=1$
is to add enough systematic uncertainty in quadrature to the statistical
uncertainty so this total uncertainty is equal to the absolute value of the
mean difference.  This is only performed when the 
absolute value of the mean difference is greater than the 
statistical uncertainty.  Following this second approach and assuming
that the stars in common are representative of the whole samples suggests
that LOSS is systematically brighter than CfA4 by 0.008 mag in $B$ and 
fainter by 0.011 mag in $V$.  LOSS is 0.034 mag brighter
than CSP2 in $B$ and consistent in $V$.  CSP2 would be be 0.022 mag
fainter than CfA4 in $B$ and 0.015 mag fainter in $V$.  However, it should
be noted that there are only $\sim$80 stars in common between LOSS and CfA4
and between CSP2 and CfA4, and only $\sim$45 between LOSS and CSP2.
Histograms that show all of the comparison star differences are presented in 
Figures \ref{fig_loss-cfa4_star_diff},
\ref{fig_csp2-cfa4_star_diff}, and \ref{fig_loss-csp2_star_diff}.  The 
distributions are reasonably symmetric around their mean except for the
$B$ LOSS-CSP2 histogram.

In order to compare the SN photometry a cubic spline was fit to the light curve 
from one group (for descriptive purposes here, group A).  The spline was 
always of the same order.  It was
allowed to extend one day beyond the earliest and latest 
points.  It was visually inspected to ensure that it smoothly fit the
data.  This spline was then
subtracted from the other group's points (group B) that were within 
the spline's date range
and had at least one data point from group A within four days.  For each SN
that we compared we fit a spline to each group's photometry and 
subtracted the other group's points.  In cases where the light curves of
each group are
roughly equally well-sampled and smooth there is virtually no difference
between which group's data is used for the spline.  In these cases we
presented the subtraction direction that gave rise to a slightly
smaller reduced $\chi^2$.  In cases where one group's
light curve is more densely sampled and/or smooth a superior
spline fit was produced and we used that one to perform the
comparison.  In light of
the inherent limitations of comparing non-s-corrected photometry, which only
allows for a reasonable comparison, we opted not to do the slightly more
involved task of combining both subtraction directions.

\begin{figure}
\includegraphics[scale=.80]{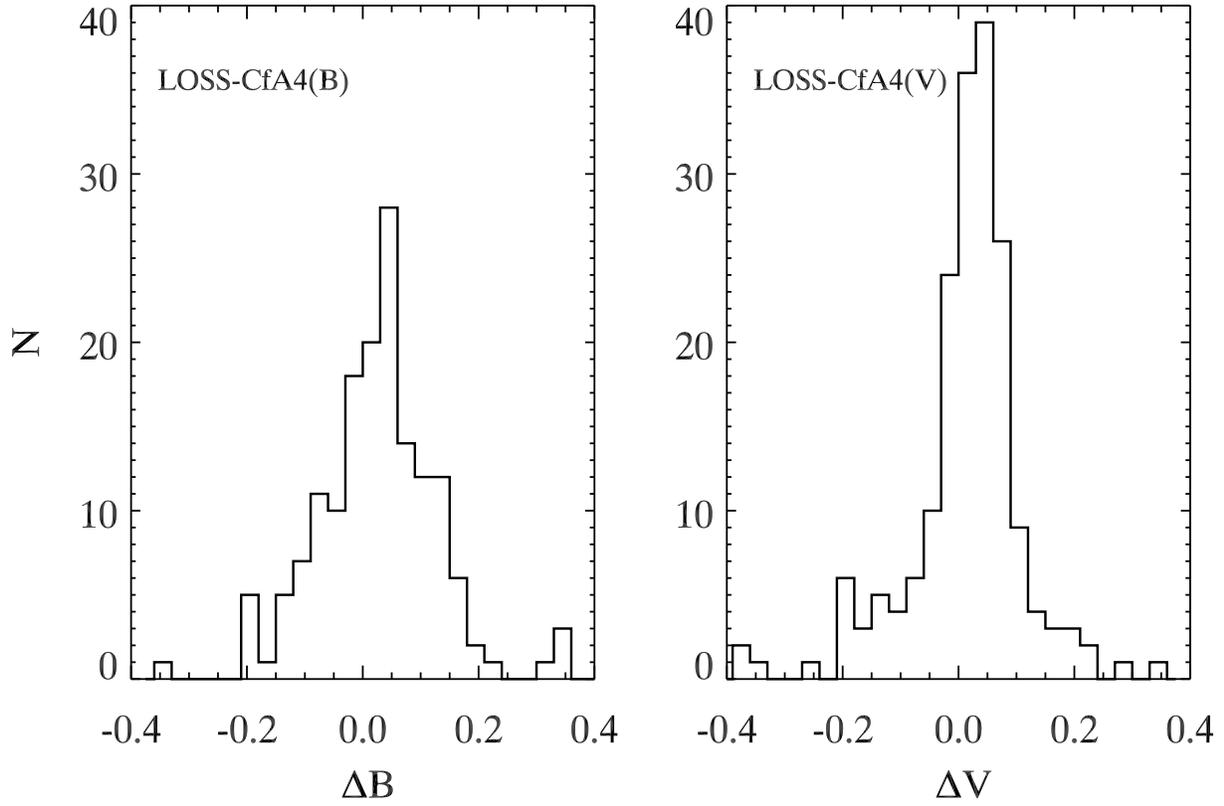}
\caption{Histograms of the $BV$ differences of the LOSS-CfA4 SN 
photometry for the 12 SN in common between the two samples.  The 
distributions are roughly symmetric.}
\label{fig_loss-cfa4_diff}
\end{figure}

\begin{figure}
\vspace{3.0in}
\scalebox{1.10}[1.10]{
\plotone{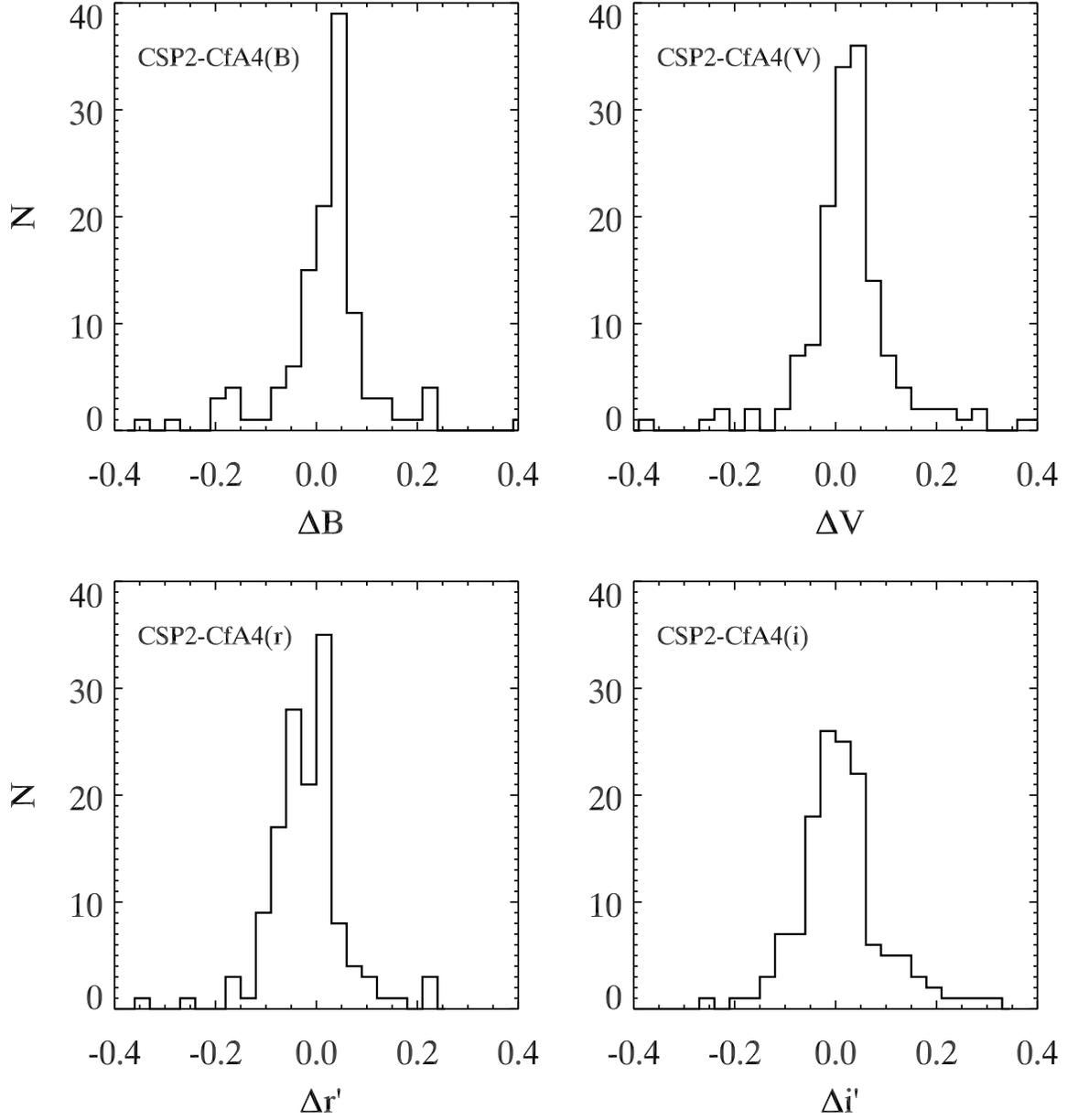}
}
\caption{Histograms of the $BVr'i'$ differences of the CSP2-CfA4 SN photometry 
for the eight SN in common between the two samples.  The distributions are 
roughly symmetric.}
\label{fig_csp2-cfa4_diff}
\end{figure}

The weighted mean of the differences of all of the subtracted points between
two groups in a given band is presented in Table 
\ref{table_complc_sum}.  There is general agreement between CfA4, LOSS and CSP2
in these weighted means.
CfA4 is $\sim0.03$ mag brighter than CSP2 and LOSS in $BV$, 
$\sim0.02$ mag fainter than CSP2 in $r'$ and $0.004$ mag brighter than
CSP2 in $i'$.  The reduced $\chi^2$ ranges from 1.8 to 3.6.  These should
be understood as a modified reduced $\chi^2$ since the spline's
degrees of freedom were not included in the calculation.  The standard
deviation of the differences is $\sim$0.1 mag.  This is reduced to 
$0.04-0.05$ mag for the CSP2-CfA4 comparison when points fainter than 18 mag 
are removed and to $0.07$ mag for LOSS-CfA4.  A comparison of LOSS versus CSP 
objects was only performed on the three SN that were also in common with CfA4.  
Histograms of the 
LOSS-CfA4 and CSP2-CfA4 differences for all the subtracted points are shown
in each filter in Figures \ref{fig_loss-cfa4_diff} and 
\ref{fig_csp2-cfa4_diff}, respectively.  The distributions of the
differences are all roughly symmetric.  The lack of s-corrections, to take
into account the response of different detectors to SN Ia SEDs,
makes the SN photometry comparisons less accurate but still give a reasonable
estimate of how well they agree.  Accurate s-corrections may resolve
some of the discrepancies.

Table \ref{table_compstars_lc} presents the comparisons of the individual
SN, showing the mean difference and number of comparison stars in common, as 
well as the weighted mean and reduced $\chi^2$ of the SN photometry 
differences.  There seems to be very little correlation between the 
differences in the comparison star and SN photometry between LOSS and CfA4 
but there does seem to be between CSP2 and CfA4.  

The comparisons of the three SN Ia in common between CfA4, LOSS and CSP2
are shown in Table \ref{table_3commonsn}.  For SN 2007le, the best agreement
in $V$ is between LOSS and CfA4 and in $B$ is between LOSS and CSP2.  For
SN 2008C, the best agreement in $V$ is between LOSS and CSP2 and in $B$ is between
LOSS and CfA4.  Finally, for SN 2009dc, the best agreement in $V$ is
between CSP2 and CfA4 and in $B$ is between LOSS and CSP2.

The main message from these comparisons is that the three groups are in
reasonable agreement but that systematic uncertainties and effects are 
present in both the comparison stars and the SN photometry.  A more definitive
comparison of the SN photometry would require accurate s-corrections.

The primary goal of SN Ia photometry is to produce accurate distances for
cosmological purposes.  \citet{mandel12} will provide an in-depth analysis
of nearby distances that will include all recently-published nearby optical and
NIR photometry.  Part of this will examine offsets between different
samples.

\section{Conclusion }

The CfA4 sample consists of 94 nearby SN Ia optical light curves.  Most of these
are new objects while 17 of them were also observed by LOSS or CSP2 and have adequate
agreement between the different groups.  The CfA4 sample is presented in both
standard and natural systems.  Each of our 94 SN Ia data images had at least 
one reference images subtracted.  In most cases, we had multiple reference
images, leading to improved knowledge of the net flux and of its uncertainty.
CfA4 is the first large nearby 
sample to have its natural-system passbands determined by use of a tunable laser 
and calibrated photodiode \citep{cramer12}.  Deposits/condensation on the 
camera likely caused there to be two time periods with different average 
$B-V$ and $V-r'$ color coefficients and natural
system passbands.  However, the separation of the photometry, calibration
and natural-system passbands into the two time periods takes care of this problem.

Systematic uncertainties are now the largest obstacle in improving understanding
of the expansion history of the universe.  One of these systematic uncertainties
is in the SN Ia photometry itself.  Ensuring stable instruments and
understanding the detector passbands involved is critical.  In the case of CfA4
the deposits/condensation shifted the passbands but careful calibrations--both the
standard star observations and the C12 passband measurements--enabled this
to be understood and overcome.

\acknowledgments

We thank the staff at FLWO for their dedicated work in maintaining the 1.2m
telescope and instruments.  We also thank M. Stritzinger, W. Li, and M. 
Ganeshalingam for help in comparing the CfA4 sample with the CSP2 and
LOSS samples.  Finally, we appreciate discussions with K. Mandel.
This work has been supported, in part, by NSF grant AST0606772 
and AST0907903 to Harvard University.  

\facility{\emph{Facility}:  FLWO:1.2m}


\clearpage
\begin{deluxetable}{llllllll}
\rotate
\tabletypesize{\scriptsize}
\tablecolumns{8}
\tablewidth{0pc}
\tablecaption{SN Ia Discovery Data}
\tablecomments{
J2000 positions are calibrated against UCAC3 in all but a few cases where 
there was insufficient coverage and USNO-A2.0 or  USNO-B1.0 was used instead.
The positions are a weighted mean of our measured SN RA and DEC, usually from 
$r'$ but occasionally from $V$ when insufficient $r'$ data were available.  
These are usually an improvement over the positions reported by the
discoverer.  The Galaxy column lists the cross-identification object from NED
with an underscore replacing any spaces in the name to facilitate the table's use 
in a columnated format.  
The redshifts, $z_{\rm helio}$ and $z_{\rm CMB}$, are primarily from NED with a few 
coming from IAUC/CBET/ATEL sources when none were available from NED.  Milky Way
E(B-V) values are taken from the NASA IPAC webform:  
http://irsa.ipac.caltech.edu/applications/DUST/ except for SN 2007fq and SN 
2009hp which are from the \citet{schlegel98} values provided by NED.  Finally, the SN
discovery reference is listed.  The majority are from CBET, IAUC or ATEL while two 
SN come from the SNF website:  http://snfactory.lbl.gov/snf/open\_access/snlist.php.
}
\tablehead{ \colhead{SN Ia} & \colhead{Position} & \colhead{Galaxy} & \colhead{$z_{\rm helio}$} & \colhead{$z_{\rm CMB}$} & \colhead{$E(B-V)$} & \colhead{$dE(B-V)$} & \colhead{Disc. Ref.} 
}
\startdata
2006ct & 12:09:56.851 +47:05:44.31 & 2MASX\_J12095669+4705461 & 0.0315 & 0.0322 & 0.0191 & 0.0014 & IAUC 8720 \\
2006ou & 11:37:13.039 +15:26:06.59 & UGC\_6588 & 0.0135 & 0.0146 & 0.0334 & 0.0016 & IAUC 8781 \\
2007A  & 00:25:16.681 +12:53:12.78 & NGC\_105 & 0.0177 & 0.0165 & 0.0736 & 0.0019 & CBET  795 \\
2007aj & 12:47:54.524 +54:00:38.08 & CGCG\_270-24 & 0.0110 & 0.0115 & 0.0163 & 0.0016 & IAUC 8822 \\
2007bj & 16:22:10.589 -01:30:51.33 & NGC\_6172 & 0.0167 & 0.0170 & 0.1177 & 0.0029 & IAUC 8834 \\
2007cb & 13:58:17.199 -23:22:21.68 & ESO\_510-G31 & 0.0366 & 0.0375 & 0.0719 & 0.0011 & IAUC 8843 \\
2007cc & 14:08:42.050 -21:35:47.50 & ESO\_578-G26 & 0.0291 & 0.0300 & 0.0794 & 0.0030 & IAUC 8843 \\
2007cf & 15:23:07.676 +08:31:45.79 & CGCG\_77-100 & 0.0329 & 0.0335 & 0.0343 & 0.0007 & IAUC 8843 \\
2007cn & 22:13:55.790 +13:45:23.45 & UGC\_11953 & 0.0253 & 0.0241 & 0.0621 & 0.0003 & IAUC 8851 \\
2007cs & 23:49:38.930 +29:55:52.61 & UGC\_12798 & 0.0176 & 0.0164 & 0.0662 & 0.0022 & CBET  986 \\
2007ev & 22:40:06.201 +24:41:56.67 & AGC\_320702 & 0.0427 & 0.0416 & 0.0490 & 0.0006 & CBET  991 \\
2007fb & 23:56:52.383 +05:30:31.90 & UGC\_12859 & 0.0180 & 0.0168 & 0.0556 & 0.0011 & IAUC 8864 \\
2007fq & 20:34:55.742 -23:06:15.38 & MCG\_-04-48-019 & 0.0425 & 0.0416 & 0.0420 & ... & CBET 1001 \\
2007fs & 22:01:40.450 -21:30:30.22 & ESO\_601-G5 & 0.0172 & 0.0162 & 0.0336 & 0.0007 & IAUC 8864 \\
2007hg & 04:08:32.676 +02:22:43.20 & [ISI96]\_0405+0214 & 0.0300 & 0.0297 & 0.3799 & 0.0106 & CBET 1047 \\
2007hj & 23:01:47.880 +15:35:11.23 & NGC\_7461 & 0.0141 & 0.0129 & 0.0883 & 0.0120 & IAUC 8874 \\
2007hu & 16:56:29.887 +27:58:39.75 & NGC\_6261 & 0.0354 & 0.0354 & 0.0458 & 0.0023 & CBET 1056 \\
2007if & 01:10:51.370 +15:27:39.63 & [YQ2007]\_J011051.37+152739.9 & 0.0742 & 0.0731 & 0.0831 & 0.0066 & CBET 1059 \\
2007ir & 02:33:41.898 +37:40:08.12 & UGC\_2033 & 0.0352 & 0.0345 & 0.0495 & 0.0009 & CBET 1067 \\
2007is & 16:47:14.607 +40:14:36.40 & UGC\_10553 & 0.0297 & 0.0297 & 0.0201 & 0.0012 & IAUC 8874 \\
2007jg & 03:29:50.815 +00:03:24.55 & SDSS\_J032950.83+000316.0 & 0.0371 & 0.0366 & 0.1065 & 0.0023 & CBET 1076 \\
2007kd & 09:25:58.041 +34:38:00.11 & MCG\_+06-21-36 & 0.0242 & 0.0250 & 0.0217 & 0.0005 & IAUC 8874 \\
2007kf & 17:31:31.266 +69:18:39.59 & [K2007]J173131.76+691840.1 & 0.0460 & 0.0458 & 0.0439 & 0.0011 & IAUC 8875 \\
2007kg & 23:58:37.493 +60:59:07.41 & 2MFGC\_18005 & 0.0070 & 0.0063 & 0.9977 & 0.0238 & IAUC 8875 \\
2007kh & 03:15:12.049 +43:10:13.39 & [YAA2007a]J031512.10+431013.0 & 0.0500 & 0.0495 & 0.1984 & 0.0030 & CBET 1089 \\
2007kk & 03:42:23.258 +39:14:30.30 & UGC\_2828 & 0.0410 & 0.0406 & 0.2291 & 0.0132 & CBET 1096 \\
2007le & 23:38:48.452 -06:31:21.83 & NGC\_7721 & 0.0067 & 0.0055 & 0.0334 & 0.0003 & CBET 1100 \\
2007nq & 00:57:33.721 -01:23:20.29 & UGC\_595 & 0.0450 & 0.0439 & 0.0354 & 0.0012 & CBET 1106 \\
2007ob & 23:12:25.988 +13:54:49.13 & 2MASX\_J23122598+1354503 & 0.0339 & 0.0327 & 0.0681 & 0.0015 & CBET 1112 \\
2007rx & 23:40:11.782 +27:25:15.59 & BATC\_J234012.05+272512.23 & 0.0301 & 0.0289 & 0.0890 & 0.0078 & CBET 1157 \\
2007ss & 12:41:06.150 +50:23:28.51 & NGC\_4617 & 0.0155 & 0.0161 & 0.0149 & 0.0004 & CBET 1175 \\
2007su & 22:19:08.884 +13:10:39.89 & SDSS\_J221908.85+131040.5 & 0.0279 & 0.0267 & 0.0830 & 0.0006 & CBET 1178 \\
2007sw & 12:13:36.933 +46:29:36.56 & UGC\_7228 & 0.0252 & 0.0260 & 0.0186 & 0.0014 & CBET 1185 \\
2007ux & 10:09:19.939 +14:59:33.07 & 2MASX\_J10091969+1459268 & 0.0309 & 0.0320 & 0.0448 & 0.0008 & CBET 1187 \\
2008A  & 01:38:17.394 +35:22:13.06 & NGC\_634 & 0.0165 & 0.0156 & 0.0542 & 0.0024 & CBET 1193 \\
2008C  & 06:57:11.469 +20:26:13.58 & UGC\_3611 & 0.0166 & 0.0171 & 0.0839 & 0.0026 & CBET 1195 \\
2008Q  & 01:24:57.207 +09:33:01.30 & NGC\_524 & 0.0600 & 0.0590 & 0.0828 & 0.0010 & CBET 1228 \\
2008Y  & 11:19:30.581 +54:27:46.21 & MCG\_+9-19-39 & 0.0697 & 0.0703 & 0.0129 & 0.0009 & CBET 1240 \\
2008Z  & 09:43:15.258 +36:17:03.64 & SDSS\_J094315.36+361709.2 & 0.0210 & 0.0218 & 0.0114 & 0.0007 & CBET 1243 \\
2008ac & 11:53:45.200 +48:25:20.79 & SDSS\_J115345.22+482521.0 & 0.0528 & 0.0535 & 0.0190 & 0.0003 & CBET 1245 \\
2008ad & 12:49:37.071 +28:19:45.82 & ROTSE\_J124936.88+281944.8 & 0.0500 & 0.0509 & 0.0130 & 0.0010 & CBET 1245 \\
2008ae & 09:56:03.160 +10:29:58.52 & IC\_577 & 0.0301 & 0.0312 & 0.0277 & 0.0007 & CBET 1247 \\
2008ai & 10:57:39.957 +37:39:41.40 & CGCG\_184-39 & 0.0353 & 0.0361 & 0.0163 & 0.0010 & CBET 1256 \\
2008ar & 12:24:37.922 +10:50:16.74 & IC\_3284 & 0.0261 & 0.0272 & 0.0373 & 0.0013 & CBET 1273 \\
2008at & 10:27:12.469 +71:24:55.55 & UGC05645 & 0.0350 & 0.0352 & 0.0912 & 0.0025 & CBET 1277 \\
2008bi & 08:35:53.388 +00:42:22.85 & NGC\_2618 & 0.0134 & 0.0144 & 0.0441 & 0.0014 & CBET 1312 \\
2008bw & 18:26:50.440 +51:08:16.42 & UGC\_11241 & 0.0331 & 0.0328 & 0.0399 & 0.0017 & CBET 1346 \\
2008by & 12:05:20.907 +40:56:44.43 & SDSS\_J120520.81+405644.4 & 0.0450 & 0.0458 & 0.0135 & 0.0002 & CBET 1350 \\
2008bz & 12:38:57.686 +11:07:45.60 & 2MASX\_J12385810+1107502 & 0.0603 & 0.0614 & 0.0269 & 0.0021 & CBET 1353 \\
2008cd & 13:15:01.777 -15:57:06.70 & NGC\_5038 & 0.0074 & 0.0085 & 0.0688 & 0.0002 & CBET 1360 \\
2008cf & 14:07:32.585 -26:33:07.74 & [WLF2008]\_J140732.38-263305.6 & 0.0460 & 0.0469 & 0.0674 & 0.0017 & CBET 1365 \\
2008cm & 13:29:12.826 +11:16:20.65 & NGC\_2369 & 0.0111 & 0.0116 & 0.1139 & 0.0013 & CBET 1384 \\
2008dr & 22:10:51.664 +02:06:29.34 & NGC\_7222 & 0.0414 & 0.0403 & 0.0428 & 0.0009 & CBET 1419 \\
2008ds & 00:29:50.820 +31:23:33.88 & UGC\_299 & 0.0210 & 0.0200 & 0.0643 & 0.0028 & CBET 1419 \\
2008dt & 16:56:30.592 +27:58:33.83 & NGC\_6261 & 0.0354 & 0.0354 & 0.0458 & 0.0021 & CBET 1423 \\
2008fr & 01:11:49.224 +14:38:26.21 & SDSS\_J011149.19+143826.5 & 0.0490 & 0.0479 & 0.0449 & 0.0014 & CBET 1513 \\
2008gb & 02:57:57.141 +46:51:56.19 & UGC\_2427 & 0.0370 & 0.0364 & 0.1983 & 0.0041 & CBET 1527 \\
2008gl & 01:20:54.820 +04:48:19.22 & UGC\_881 & 0.0340 & 0.0330 & 0.0284 & 0.0009 & CBET 1545 \\
2008hj & 00:04:01.913 -11:10:08.35 & MCG\_-2-1-14 & 0.0379 & 0.0367 & 0.0361 & 0.0009 & CBET 1579 \\
2008hm & 03:27:10.889 +46:56:39.20 & 2MFGC\_2845 & 0.0197 & 0.0192 & 0.4425 & 0.0099 & CBET 1586 \\
2008hs & 02:25:29.594 +41:50:34.92 & NGC\_910 & 0.0173 & 0.0166 & 0.0573 & 0.0004 & CBET 1598 \\
2008hv & 09:07:34.066 +03:23:32.18 & NGC\_2765 & 0.0125 & 0.0136 & 0.0321 & 0.0009 & CBET 1601 \\
2009D  & 03:54:22.817 -19:10:54.56 & MCG\_-03-10-52 & 0.0250 & 0.0247 & 0.0529 & 0.0014 & CBET 1647 \\
2009Y  & 14:42:24.563 -17:14:46.70 & NGC\_5728 & 0.0093 & 0.0101 & 0.1016 & 0.0006 & CBET 1684 \\
2009ad & 05:03:33.393 +06:39:35.82 & UGC\_3236 & 0.0284 & 0.0283 & 0.1120 & 0.0013 & CBET 1694 \\
2009al & 10:51:22.049 +08:34:41.98 & NGC\_3425 & 0.0221 & 0.0233 & 0.0246 & 0.0005 & CBET 1705 \\
2009an & 12:22:47.385 +65:51:04.60 & NGC\_4332 & 0.0092 & 0.0095 & 0.0186 & 0.0003 & CBET 1707 \\
2009bv & 13:07:20.517 +35:47:03.20 & MCG\_+6-29-39 & 0.0367 & 0.0375 & 0.0086 & 0.0009 & CBET 1741 \\
2009dc & 15:51:12.083 +25:42:28.43 & UGC\_10064 & 0.0214 & 0.0217 & 0.0696 & 0.0017 & CBET 1762 \\
2009do & 12:34:58.316 +50:51:03.81 & NGC\_4537 & 0.0397 & 0.0403 & 0.0149 & 0.0006 & CBET 1778 \\
2009ds & 11:49:04.025 -09:43:44.48 & NGC\_3905 & 0.0192 & 0.0204 & 0.0389 & 0.0007 & CBET 1784 \\
2009fv & 16:29:44.191 +40:48:41.44 & NGC\_6173 & 0.0293 & 0.0294 & 0.0063 & 0.0014 & CBET 1834 \\
2009gf & 14:15:37.127 +14:16:48.74 & NGC\_5525 & 0.0185 & 0.0193 & 0.0255 & 0.0007 & CBET 1844 \\
2009hp & 02:58:23.938 +06:35:34.64 & MCG\_+01-08-30 & 0.0211 & 0.0204 & 0.2300 & ... & CBET 1888 \\
2009ig & 02:38:11.613 -01:18:45.52 & NGC\_1015 & 0.0088 & 0.0080 & 0.0320 & 0.0009 & CBET 1918 \\
2009jr & 20:26:26.013 +02:54:31.73 & IC\_1320 & 0.0165 & 0.0156 & 0.1347 & 0.0035 & CBET 1964 \\
2009kk & 03:49:44.320 -03:15:52.66 & 2MFGC\_3182 & 0.0129 & 0.0124 & 0.1376 & 0.0029 & CBET 1991 \\
2009kq & 08:36:15.148 +28:04:01.67 & MCG\_+5-21-1 & 0.0116 & 0.0124 & 0.0410 & 0.0006 & CBET 2005 \\
2009le & 02:09:17.160 -23:24:44.74 & ESO\_478-6 & 0.0178 & 0.0170 & 0.0164 & 0.0006 & CBET 2022 \\
2009lf & 02:01:39.616 +15:19:58.13 & 2MASX\_J02014081+1519521 & 0.0450 & 0.0441 & 0.0525 & 0.0023 & CBET 2023 \\
2009li & 00:22:51.395 +06:58:11.35 & IC\_1549 & 0.0404 & 0.0392 & 0.0267 & 0.0010 & CBET 2026 \\
2009na & 10:47:01.444 +26:32:37.73 & UGC\_5884 & 0.0210 & 0.0220 & 0.0319 & 0.0018 & CBET 2098 \\
2009nq & 23:15:17.004 +19:01:21.58 & NGC\_7549 & 0.0158 & 0.0146 & 0.1455 & 0.0046 & CBET 2110 \\
2010A  & 02:32:39.459 +00:37:09.90 & UGC\_2019 & 0.0207 & 0.0199 & 0.0291 & 0.0011 & CBET 2109 \\
2010H  & 08:06:24.342 +01:02:09.01 & IC\_494 & 0.0152 & 0.0160 & 0.0308 & 0.0011 & CBET 2130 \\
2010Y  & 10:51:03.994 +65:46:46.40 & NGC\_3392 & 0.0109 & 0.0113 & 0.0135 & 0.0015 & CBET 2168 \\
2010ag & 17:03:53.653 +31:30:06.70 & UGC\_10679 & 0.0338 & 0.0338 & 0.0309 & 0.0013 & CBET 2195 \\
2010ai & 12:59:24.005 +27:59:47.13 & SDSS\_J125925.04+275948.2 & 0.0184 & 0.0193 & 0.0094 & 0.0012 & CBET 2200 \\
2010cr & 13:29:25.082 +11:47:46.49 & NGC\_5177 & 0.0216 & 0.0225 & 0.0345 & 0.0014 & CBET 2281 \\
2010dt & 16:43:15.063 +32:40:27.56 & CGCG\_168-029 & 0.0529 & 0.0529 & 0.0341 & 0.0011 & CBET 2307 \\
2010dw & 15:22:40.279 -05:55:16.46 & 2MASX\_J15224062-0555214 & 0.0381 & 0.0387 & 0.0933 & 0.0011 & CBET 2310 \\
SNF20080522-000 & 13:36:47.592 +05:08:30.41 & SDSS\_J133647.59+050833.0 & 0.0472 & 0.0482 & 0.0265 & 0.0005 & SNF site \\
SNF20080522-011 & 15:19:58.920 +04:54:16.73 & SDSS\_J151959.16+045411.2 & 0.0397 & 0.0403 & 0.0427 & 0.0008 & SNF site \\
PTF10bjs & 13:01:11.215 +53:48:57.49 & MCG\_+9-21-83 & 0.0300 & 0.0306 & 0.0176 & 0.003 & ATEL 2453 \\
\enddata
\label{table_sndiscovery}
\end{deluxetable}

\clearpage
\begin{deluxetable}{lrcc}
\tabletypesize{\scriptsize}
\tablecolumns{4}
\tablewidth{0pc}
\tablecaption{Photometric Color Terms}
\tablecomments{
~Lower-case $ubvri$ refer to the instrumental magnitudes while
\emph{Uu'BVr'i'} refer to the standard magnitudes.  All color terms
implicitly contain an additive constant.  For example, 
$(v-V) = 0.0233(B-V)$ + const, $(v-i) = 1.0239(V-i')$ + const.
}
\tablehead{
\colhead{Filter/Time Period} & \colhead{Color Term} &  \colhead{Value} & \colhead{Nights}
}
\startdata
$U-B$/both periods  & $(u-b)/(U-B)$  &  $0.9981 \pm 0.0209$ & 17\\
$u'-B$/both periods & $(u-b)/(u'-B)$ &  $0.9089 \pm 0.0057$ & 28\\
$B-V$/period one& $(b-v)/(B-V)$  &  $0.9294 \pm 0.0026$ & 38\\
$B-V$/period two& $(b-v)/(B-V)$  &  $0.8734 \pm 0.0024$ & 25\\
$V$/both periods  & $(v-V)/(B-V)$  &  $0.0233 \pm 0.0018$ & 63\\
$V-r'$/period one& $(v-r)/(V-r')$ &  $1.0684 \pm 0.0028$ & 38\\
$V-r'$/period two& $(v-r)/(V-r')$ &  $1.0265 \pm 0.0033$ & 25\\
$V-i'$/both periods& $(v-i)/(V-i')$ &  $1.0239 \pm 0.0016$ & 63\\
\enddata
\label{table_colorterms}
\end{deluxetable}

\clearpage
\begin{deluxetable}{lcccccc}
\tabletypesize{\scriptsize}
\tablecolumns{7}
\tablewidth{0pc}
\tablecaption{Estimation of $\sigma_{\rm phot}$ for Single Reference-Image Photometry}
\tablecomments{This estimate of $\sigma_{\rm phot}$ for the $N=1$ cases
was based on a quadratic fit of the standard deviations of the
multiple-subtraction photometry values versus SN magnitude in a given band at
magnitudes fainter than 16 mag.  For magnitudes brighter than this a constant
but representative value was used.  The CfA3 photometry uncertainties would
be improved by adding these values in quadrature.}
\tablehead{
\colhead{Mag} & \colhead{$U$} &  \colhead{$u'$} & \colhead{$B$} & \colhead{$V$} &  \colhead{$r'$} &  \colhead{$i'$}
}
\startdata
$\leq$16.0 & 0.020 & 0.020 & 0.012 & 0.010 & 0.010 & 0.010\\
16.5 & 0.030 & 0.038 & 0.016 & 0.013 & 0.012 & 0.013\\
17.0 & 0.048 & 0.058 & 0.024 & 0.019 & 0.018 & 0.019\\
17.5 & 0.073 & 0.081 & 0.037 & 0.028 & 0.026 & 0.028\\
18.0 & 0.105 & 0.107 & 0.053 & 0.041 & 0.037 & 0.041\\
18.5 & 0.144 & 0.136 & 0.074 & 0.058 & 0.052 & 0.056\\
19.0 & 0.191 & 0.168 & 0.099 & 0.077 & 0.070 & 0.075\\
19.5 & 0.245 & 0.203 & 0.128 & 0.100 & 0.091 & 0.097\\
20.0 & 0.306 & 0.240 & 0.162 & 0.126 & 0.115 & 0.122\\
20.5 & 0.374 & 0.281 & 0.199 & 0.156 & 0.142 & 0.150\\
21.0 & 0.450 & 0.324 & 0.241 & 0.189 & 0.172 & 0.181\\
\enddata
\label{table_singlerefimage}
\end{deluxetable}

\begin{deluxetable}{ccccccccccccccccccc}
\rotate
\tabletypesize{\tiny}
\tablecolumns{19}
\tablewidth{0pc}
\tablecaption{Standard System Comparison Star Photometry}
\tablecomments{This table presents the CfA4 standard system 
comparison star photometry.  Only a portion of the table
is shown here.  The complete table is available from the online journal
or from the CfA website.  The period-one and two natural
system values can be calculated by applying the color terms or
are available upon request.  All $u'-B$ comparison star photometry 
is presented here and in the complete online version as $U-B$ but can 
readily be converted to $u'-B$ via the equation $u'=U+0.854$ \citep{chonis08}.
The $u'$ SN photometry is presented as $u'$ in Tables \ref{table_lc_nat} and
\ref{table_lc_std}.
}
\tablehead{
\colhead{SN} & \colhead{Star} & \colhead{RA} & \colhead{DEC} & \colhead{$V$} & \colhead{$\sigma$} & \colhead{$N$} 
& \colhead{$U-B$} & \colhead{$\sigma$} & \colhead{$N$} & \colhead{$B-V$} & \colhead{$\sigma$} & \colhead{$N$}
& \colhead{$V-r'$} & \colhead{$\sigma$} & \colhead{$N$} & \colhead{$V-i'$} & \colhead{$\sigma$} & \colhead{$N$}
}
\startdata
2010ai & 01&12:59:44.963&+27:56:57.54& 15.365& 0.015&3 & ~~...~ & ... &0  & 0.837& 0.021&3 & ~0.271& 0.010&3 & ~0.477& 0.015&3\\
2010ai & 02&12:59:41.382&+28:00:07.92& 16.715& 0.015&7 & ~0.115 &0.131&1  & 0.620& 0.019&7 & ~0.185& 0.010&7 & ~0.345& 0.012&7\\
2010ai & 03&12:59:34.571&+27:54:52.44& 17.505& 0.013&7 & -0.823 &0.174&1  & 0.077& 0.020&7 & -0.088& 0.016&7 & -0.281& 0.021&7\\
2010ai & 04&12:59:28.781&+27:56:15.51& 12.449& 0.012&7 & ~~...~ & ... &0  & 1.152& 0.016&7 & ~~...~&  ... &0 & ~~...~&  ... &0\\
2010ai & 05&12:59:26.528&+28:00:24.01& 16.656& 0.017&7 & ~0.014 &0.129&1  & 0.639& 0.019&7 & ~0.157& 0.013&7 & ~0.283& 0.023&7\\
2010ai & 06&12:59:25.270&+27:59:07.64& 17.583& 0.015&7 & ~~...~ & ... &0  & 0.527& 0.029&7 & ~0.145& 0.012&7 & ~0.276& 0.030&7\\
2010ai & 07&12:59:24.761&+27:56:24.16& 15.921& 0.013&7 & ~~...~ & ... &0  & 0.950& 0.016&7 & ~0.287& 0.009&7 & ~0.474& 0.010&7\\
2010ai & 08&12:59:18.650&+28:01:43.34& 15.433& 0.012&7 & ~~...~ & ... &0  & 1.260& 0.016&7 & ~0.497& 0.009&7 & ~0.909& 0.010&7\\
2010ai & 09&12:59:15.882&+27:57:10.68& 16.109& 0.013&7 & -0.200 &0.110&1  & 0.523& 0.018&7 & ~0.113& 0.009&7 & ~0.202& 0.013&7\\
2010ai & 10&12:59:13.739&+28:02:10.48& 16.916& 0.013&7 & -0.161 &0.141&1  & 0.522& 0.018&7 & ~0.124& 0.011&7 & ~0.249& 0.011&7\\
2010ai & 11&12:59:11.788&+28:00:04.02& 15.492& 0.012&7 & ~0.302 &0.096&1  & 0.784& 0.016&7 & ~0.234& 0.009&7 & ~0.398& 0.010&7\\
2010ai & 12&12:59:04.168&+28:03:48.67& 16.084& 0.012&7 & ~0.250 &0.110&1  & 0.761& 0.018&7 & ~0.214& 0.010&7 & ~0.389& 0.012&7\\
2010ai & 13&12:59:01.189&+28:02:03.98& 15.265& 0.013&7 & ~0.416 &0.091&1  & 0.941& 0.017&7 & ~0.293& 0.010&7 & ~0.557& 0.010&7\\ 
\enddata
\label{table_star_std}
\end{deluxetable}

\begin{deluxetable}{llrrrrrrr}
\tabletypesize{\scriptsize}
\tablecolumns{9}
\tablewidth{0pc}
\tablecaption{Natural System SN Photometry}
\tablecomments{This table presents the CfA4 natural system SN photometry.
The number of successful subtractions from that night
that survived outlier rejection is listed in the fourth column.  
For example, if there were two data images and seven reference images
and none were rejected then $N=14$.  Usually there was only one data 
image per night.  The median of the pipeline-generated uncertainties
of the surviving photometry points is listed in the fifth column.  The
standard deviation of the surviving photometry values for
that date is listed in the
the sixth column.  These two values are added in quadrature to produce
the total uncertainty.  The last column lists what period the photometry
belongs to and is of crucial importance so that the corresponding
period one or two passbands or color terms are used.  Period one is
before MJD=55058 and period two is after.  Only the first five nights
in each band of 
one SN are shown here.  The complete table with all bands and all SN is 
available from the online journal or from the CfA website.
}
\tablehead{
\colhead{SN} & \colhead{Filter} & \colhead{MJD} & \colhead{$N$} & \colhead{$\sigma_{\rm pipe}$} & \colhead{$\sigma_{\rm phot}$} & \colhead{Mag} & \colhead{$\sigma_{\rm total}$} & \colhead{Period}
}
\startdata
2010ai & B & 55267.40470 & 7 & 0.0240 & 0.0102 & 17.2950 & 0.0261 & two \\
2010ai & B & 55268.30587 & 5 & 0.0210 & 0.0227 & 16.9950 & 0.0309& two \\
2010ai & B & 55269.37863 & 7 & 0.0220 & 0.0131 & 16.7740 & 0.0256& two \\
2010ai & B & 55270.33043 & 7 & 0.0200 & 0.0146 & 16.5800 & 0.0248& two \\
2010ai & B & 55275.36791 & 14 & 0.0170 & 0.0185 & 16.1015 & 0.0251& two \\
2010ai & V & 55267.40100 & 7 & 0.0220 & 0.0136 & 17.3110 & 0.0258& two \\
2010ai & V & 55268.30216 & 6 & 0.0215 & 0.0157 & 17.0710 & 0.0266& two \\
2010ai & V & 55269.37493 & 7 & 0.0180 & 0.0108 & 16.8000 & 0.0210& two \\
2010ai & V & 55270.32675 & 7 & 0.0180 & 0.0074 & 16.6190 & 0.0195& two \\
2010ai & V & 55271.32120 & 7 & 0.0160 & 0.0131 & 16.4630 & 0.0207& two \\
2010ai & r' & 55267.39800 & 6 & 0.0195 & 0.0062 & 17.3215 & 0.0205& two \\
2010ai & r' & 55268.29912 & 7 & 0.0190 & 0.0098 & 17.0670 & 0.0214& two \\
2010ai & r' & 55269.37192 & 7 & 0.0190 & 0.0084 & 16.7930 & 0.0208& two \\
2010ai & r' & 55270.32374 & 7 & 0.0180 & 0.0094 & 16.6060 & 0.0203& two \\
2010ai & r' & 55271.31818 & 7 & 0.0170 & 0.0077 & 16.4580 & 0.0187& two \\
2010ai & i' & 55267.39499 & 7 & 0.0260 & 0.0448 & 17.5200 & 0.0518& two \\
2010ai & i' & 55268.29611 & 7 & 0.0260 & 0.0351 & 17.2520 & 0.0437& two \\
2010ai & i' & 55269.36892 & 7 & 0.0210 & 0.0240 & 16.9790 & 0.0319& two \\
2010ai & i' & 55270.32073 & 7 & 0.0220 & 0.0274 & 16.8430 & 0.0351& two \\
2010ai & i' & 55271.31518 & 7 & 0.0200 & 0.0279 & 16.7490 & 0.0344& two \\
\enddata
\label{table_lc_nat}
\end{deluxetable}

\begin{deluxetable}{llrrrrrr}
\tabletypesize{\scriptsize}
\tablecolumns{8}
\tablewidth{0pc}
\tablecaption{Standard System SN Photometry}
\tablecomments{This table presents the CfA4 standard system SN photometry.
The number of successful subtractions from that night
that survived outlier rejection is listed in the fourth column.  
For example, if there were two data images and seven reference images
and none were rejected then $N=14$.  Usually there was only one data 
image per night.  The median of the pipeline-generated uncertainties
of the surviving photometry points is listed in the fifth column.  The
standard deviation of the surviving photometry values for
that date is listed in the
the sixth column.  These two values are added in quadrature to produce
the total uncertainty.  Only the first five nights 
in each band of one SN are shown here.  
The complete table with all bands and all SN is available from the online 
journal or from the CfA website.
}
\tablehead{
\colhead{SN} & \colhead{Filter} & \colhead{MJD} & \colhead{$N$} & \colhead{$\sigma_{\rm pipe}$} & \colhead{$\sigma_{\rm phot}$} & \colhead{Mag} &  \colhead{$\sigma_{\rm total}$}
}
\startdata
2010ai & B & 55267.40470 & 7 & 0.0240 & 0.0102 & 17.2932 & 0.0261 \\
2010ai & B & 55268.30587 & 5 & 0.0210 & 0.0227 & 16.9861 & 0.0309 \\
2010ai & B & 55269.37863 & 7 & 0.0220 & 0.0131 & 16.7710 & 0.0256 \\
2010ai & B & 55270.33043 & 7 & 0.0200 & 0.0146 & 16.5755 & 0.0248 \\
2010ai & B & 55275.36791 & 14 & 0.0170 & 0.0185 & 16.1004 & 0.0251 \\
2010ai & V & 55267.40100 & 7 & 0.0220 & 0.0136 & 17.3114 & 0.0258 \\
2010ai & V & 55268.30216 & 6 & 0.0215 & 0.0157 & 17.0730 & 0.0266 \\
2010ai & V & 55269.37493 & 7 & 0.0180 & 0.0108 & 16.8007 & 0.0210 \\
2010ai & V & 55270.32675 & 7 & 0.0180 & 0.0074 & 16.6200 & 0.0195 \\
2010ai & V & 55271.32120 & 7 & 0.0160 & 0.0131 & 16.4644 & 0.0207 \\
2010ai & r' & 55267.39800 & 6 & 0.0195 & 0.0062 & 17.3216 & 0.0205 \\
2010ai & r' & 55268.29912 & 7 & 0.0190 & 0.0098 & 17.0691 & 0.0214 \\
2010ai & r' & 55269.37192 & 7 & 0.0190 & 0.0084 & 16.7939 & 0.0208 \\
2010ai & r' & 55270.32374 & 7 & 0.0180 & 0.0094 & 16.6074 & 0.0203 \\
2010ai & r' & 55271.31818 & 7 & 0.0170 & 0.0077 & 16.4595 & 0.0187 \\
2010ai & i' & 55267.39499 & 7 & 0.0260 & 0.0448 & 17.5155 & 0.0518 \\
2010ai & i' & 55268.29611 & 7 & 0.0260 & 0.0351 & 17.2498 & 0.0437 \\
2010ai & i' & 55269.36892 & 7 & 0.0210 & 0.0240 & 16.9755 & 0.0319 \\
2010ai & i' & 55270.32073 & 7 & 0.0220 & 0.0274 & 16.8388 & 0.0351 \\
2010ai & i' & 55271.31518 & 7 & 0.0200 & 0.0279 & 16.7437 & 0.0344 \\
\enddata
\label{table_lc_std}
\end{deluxetable}

\begin{deluxetable}{l|rrrr|rrrr}
\tabletypesize{\scriptsize}
\tablecolumns{9}
\tablewidth{0pc}
\tablecaption{Comparison Star Mean Differences}
\tablecomments{~Listed are the mean, standard error of the mean, and standard deviation of the differences in the 
photometry of all of the comparison stars in common between CfA4, LOSS and 
CSP2.  Also provided is the number of comparison stars in common.
}
\tablehead{
\colhead{SN Samples} & \colhead{$\mu_{\Delta B}$} & \colhead{$\sigma/\sqrt{N}$} &  \colhead{$\sigma_{\Delta B}$} & \colhead{$N_{\rm stars}$} & \colhead{$\mu_{\Delta V}$} & \colhead{$\sigma/\sqrt{N}$} &  \colhead{$\sigma_{\Delta V}$} & \colhead{$N_{\rm stars}$}
}
\startdata
LOSS$-$CfA4 & $-$0.0087 & 0.0037 & 0.0346 & 86 &    0.0109 &0.0028 & 0.0263 & 86\\
LOSS$-$CSP2 & $-$0.0346 & 0.0071 & 0.0489 & 48 & $-$0.0016 &0.0048 & 0.0319 & 44\\
CSP2$-$CfA4 &    0.0223 & 0.0032 & 0.0289 & 81 &    0.0149 &0.0025 & 0.0217 & 77\\
\hline
& $\mu_{\Delta r'}$ & $\sigma/\sqrt{N}$ & $\sigma_{\Delta r'}$ & $N_{\rm stars}$ & $\mu_{\Delta i'}$ & $\sigma/\sqrt{N}$ & $\sigma_{\Delta i'}$ & $N_{\rm stars}$ \\
\hline
CSP2$-$CfA4 &    0.0000 &0.0033 &  0.0300 & 79 & $-$0.0059 &0.0027 & 0.0236 & 78\\
\enddata
\label{table_compstars_sum}
\end{deluxetable}

\begin{deluxetable}{l|rccc|rccc}
\tabletypesize{\scriptsize}
\tablecolumns{9}
\tablewidth{0pc}
\tablecaption{Sample-To-Sample SN Photometry Comparisons}
\tablecomments{The weighted mean, reduced $\chi^2$ and standard deviation of 
all subtractions of the SN photometry in common between CfA4, LOSS and CSP2.  
Also, the standard deviation with points fainter than 18 mag removed.
}
\tablehead{
\colhead{SN Samples} & \colhead{$\mu_{\Delta B}$} &  \colhead{$\chi^2_\nu$(B)} & \colhead{$\sigma$(B)} & \colhead{$\sigma_{<18}$(B)} &  \colhead{$\mu_{\Delta V}$} &  \colhead{$\chi^2_\nu$(V)} & \colhead{$\sigma$(V)} & \colhead{$\sigma_{<18}$(V)}
}
\startdata
LOSS$-$CfA4 & 0.036 & 3.24 & 0.110 & 0.075 & 0.035 & 3.50 & 0.099 & 0.070 \\
CSP2$-$CfA4 & 0.027 & 1.82 & 0.126 & 0.035 & 0.026 & 3.51 & 0.095 & 0.040 \\
\hline
& \colhead{$\mu_{\Delta r'}$} &  \colhead{$\chi^2_\nu$(r')} & \colhead{$\sigma$(r')} & \colhead{$\sigma_{<18}$(r')} &  \colhead{$\mu_{\Delta i'}$} &  \colhead{$\chi^2_\nu$(i')} & \colhead{$\sigma$(i')} & \colhead{$\sigma_{<18}$(i')} \\
\hline
CSP2$-$CfA4 & $-$0.019 & 3.36 & 0.096 & 0.054 & 0.004 & 2.37 & 0.092 & 0.048 \\
\enddata
\label{table_complc_sum}
\end{deluxetable}

\begin{deluxetable}{l|rrrr|rr|rrrr|rr}
\tabletypesize{\scriptsize}
\tablecolumns{13}
\tablewidth{0pc}
\tablecaption{Comparison Star and SN Photometry Comparison}
\tablecomments{~Listed are the mean, standard error of the mean, and standard deviation 
of the differences between
comparison stars in common for each SN between
CfA4 and LOSS, CfA4 and CSP2, and CSP2 and LOSS, and the number 
of stars in common.  Also listed is the weighted mean and the
$\chi^2$ of the differences in the SN photometry.
}
\tablehead{
\colhead{Samples/SN} & \colhead{$\mu_{\Delta}$} & \colhead{$\sigma/\sqrt{N}$} &  \colhead{$\sigma_{\Delta}$} & \colhead{$N_{\rm stars}$} & \colhead{$\mu_{\Delta(wgt)}$} &  \colhead{$\chi^2_\nu$} & \colhead{$\mu_{\Delta}$} & \colhead{$\sigma/\sqrt{N}$} & \colhead{$\sigma_{\Delta}$} & \colhead{$N_{\rm stars}$} & \colhead{$\mu_{\Delta(wgt)}$} &  \colhead{$\chi^2_\nu$}
}
\startdata
LOSS-CfA4& $B_{\rm stars}$&   &  &  &  $B_{\rm SN}$&     &  $V_{\rm stars}$& & & & $V_{\rm SN}$ &   \\
\hline
2007bj & 0.014&0.009&0.028&10& 0.0267& 2.126 &  0.019&0.007&0.021&10& 0.0500&5.996 \\
2007hj & 0.012&0.024&0.058&6 & 0.0438& 3.676 &  0.019&0.014&0.035&6 & 0.0438&3.550 \\
2007le &-0.061&0.050&0.071&2 & 0.0172& 3.127 & -0.032&0.052&0.074&2 & 0.0258&2.814 \\
2007ux & 0.018&0.007&0.016&5 &-0.0788& 1.454 &  0.014&0.008&0.018&5 & 0.0451&3.148 \\
2008A  &-0.021&0.011&0.043&14& 0.0568& 2.510 &  0.009&0.007&0.028&14& 0.0367&7.460 \\
2008C  &-0.022&0.006&0.027&23&-0.0133& 0.995 &  0.017&0.004&0.021&23& 0.0295&1.184 \\
2008Q  &-0.028&0.006&0.011&3 & 0.1327&55.403 & -0.046&0.005&0.008&3 &-0.0013&0.408 \\
2008Z  & 0.008&0.007&0.016&5 &-0.0173& 0.659 &  0.012&0.008&0.019&5 & 0.0278&5.087 \\
2008ar &-0.002&0.004&0.005&2 & 0.0537& 3.690 & -0.010&0.003&0.004&2 & 0.0126&4.183 \\
2008dr &-0.014&0.006&0.011&3 & 0.1129& 0.647 &  0.034&0.006&0.011&3 & 0.0115&0.527 \\
2008dt &-0.001&0.008&0.022&8 &-0.0912& 0.565 &  0.008&0.006&0.016&8 &-0.1304&4.058 \\
2008ec &-0.016&...  &...  &1 & ...   &...    & -0.009&...  &...  &1 & ...   &...   \\
2009dc &-0.002&0.007&0.015&4 & 0.0593& 3.314 &  0.013&0.007&0.014&4 & 0.0497&3.230 \\
\hline
CSP2-CfA4& $B_{\rm stars}$&  &    &  &  $B_{\rm SN}$&     &  $V_{\rm stars}$& & & & $V_{\rm SN}$ &   \\
\hline
2007A &  0.016&0.009&0.024&7 &  0.0156&2.880 & 0.005&0.005&0.011&5  & 0.0113&4.647 \\
2007if&  0.024&0.011&0.028&7 &  0.0102&0.819 & 0.017&0.005&0.014&7  & 0.0946&3.424 \\
2007jg&  0.011&0.004&0.017&15&  0.0577&1.167 &-0.004&0.005&0.020&14 & 0.0039&1.462 \\
2007le&  0.053&0.008&0.025&9 &  0.0346&4.573 & 0.042&0.006&0.017&9  & 0.0527&10.25 \\
2007nq&  0.035&0.002&0.005&5 &  0.0272&0.876 & 0.023&0.005&0.011&5  & 0.0150&1.637 \\
2008C &  0.005&0.003&0.013&15&  0.0326&2.052 & 0.028&0.004&0.014&15 & 0.0227&4.184 \\
2008hv&  0.006&0.010&0.033&12&  0.0033&1.387 & 0.005&0.006&0.020&12 &-0.0004&1.445 \\
2009dc&  0.052&0.007&0.024&11&  0.0498&2.364 & 0.011&0.007&0.021&10 & 0.0221&1.979 \\
\hline
CSP2-CfA4&$r'_{\rm stars}$&     &  & & $r'_{\rm SN}$&     & $i'_{\rm stars}$& & & & $i'_{\rm SN}$ &   \\
\hline
2007A &  0.035&0.012&0.032&7 &  0.0034&1.984 & 0.023&0.009&0.024&7  & 0.0108&0.375 \\
2007if&  0.005&0.008&0.021&7 &  0.0345&1.196 &-0.001&0.008&0.020&7  & 0.0301&2.220 \\
2007jg& -0.006&0.004&0.016&14&  0.0014&1.352 &-0.015&0.005&0.017&13 & 0.1378&3.892 \\
2007le&  0.036&0.008&0.023&9 &  0.0077&1.005 & 0.023&0.008&0.024&9  & 0.0218&2.979 \\
2007nq&  0.016&0.005&0.012&5 & -0.0286&1.056 &-0.004&0.007&0.015&5  & 0.0281&1.063 \\
2008C & -0.006&0.003&0.012&15& -0.0353&7.983 &-0.026&0.002&0.009&15 &-0.0369&4.207 \\
2008hv& -0.003&0.005&0.016&12& -0.0460&5.842 &-0.008&0.004&0.015&12 &-0.0282&3.254 \\
2009dc& -0.047&0.007&0.021&10& -0.0672&7.882 &-0.018&0.007&0.022&10 &-0.0021&0.527 \\
\hline
LOSS-CSP2& $B_{\rm stars}$&     &  &  & $B_{\rm SN}$&     &  $V_{\rm stars}$& & & & $V_{\rm SN}$ &   \\
\hline
2006bt&  0.004&0.010&0.027&7 &  ...   & ...  &  0.011&0.008&0.020&7 & ...   & ...  \\
2006ej&  0.016&...  &...  &1 &  ...   & ...  &  ...  &...  &...  &0 & ...  \\
2006hb&  0.010&0.007&0.015&5 &  ...   & ...  &  0.025&0.009&0.018&4 & ...   & ...  \\
2007af& -0.038&0.004&0.009&4 &  ...   & ...  &  0.012&0.004&0.009&4 & ...   & ...  \\
2007bc& -0.107&0.040&0.069&3 &  ...   & ...  & -0.003&0.009&0.013&2 & ...   & ...  \\
2007ca& -0.058&0.013&0.028&5 &  ...   & ...  & -0.036&0.006&0.013&5 & ...   & ...  \\
2007le& -0.109&0.032&0.064&4 &  0.0029&2.364 & -0.039&0.036&0.072&4 &-0.0343&6.137 \\
2008C & -0.018&0.008&0.025&11& -0.0198&3.314 & -0.007&0.005&0.018&11& 0.0021&1.410 \\
2009dc& -0.044&0.012&0.034&8 &  0.0030&0.903 &  0.017&0.006&0.017&7 & 0.0369&3.767 \\
\enddata
\label{table_compstars_lc}
\end{deluxetable}

\begin{deluxetable}{l|rccc|rcc}
\tabletypesize{\scriptsize}
\tablecolumns{8}
\tablewidth{0pc}
\tablecaption{Comparing the Three SN in Common}
\tablecomments{The mean difference, standard error of the mean,
 and standard deviation of the 
comparison-star photometry and number of
stars in common for each SN Ia are presented in 
columns two through five.  The
weighted mean, standard deviation and reduced $\chi^2$ of the differences
in the SN photometry are presented in the final three columns.
}
\tablehead{
\colhead{SN/Samples} & \colhead{$\mu_{\Delta}$} & \colhead{$\sigma/\sqrt{N}$} & \colhead{$\sigma_\Delta$} & \colhead{$N_{\rm stars}$} &  \colhead{$\mu_{\Delta(wgt)}$} &  \colhead{$\sigma$} & \colhead{$\chi^2_\nu$}
}
\startdata
\hline
2007le    & $B_{\rm stars}$& & & & $B_{\rm SN}$\\
CSP2-CfA4 &  0.053&0.008&0.025&9 & 0.0346 & 0.0340&4.573 \\
LOSS-CfA4 & -0.061&0.050&0.071&2 & 0.0172 & 0.0521&3.127 \\
LOSS-CSP2 & -0.109&0.032&0.064&4 & 0.0029 & 0.0434&4.204 \\
\hline
2008C     & $B_{\rm stars}$& & & & $B_{\rm SN}$\\
CSP2-CfA4 &  0.005&0.003&0.013&15& 0.0326 & 0.1976&2.052 \\
LOSS-CfA4 & -0.022&0.006&0.027&23&-0.0133 & 0.0742&0.995 \\ 
LOSS-CSP2 & -0.018&0.008&0.025&11&-0.0198 & 0.0226&1.565 \\ 
\hline
2009dc    & $B_{\rm stars}$& & & & $B_{\rm SN}$\\
CSP2-CfA4 &  0.052&0.007&0.024&11& 0.0498 & 0.0219&2.364 \\ 
LOSS-CfA4 & -0.002&0.007&0.015&4 & 0.0593 & 0.0715&3.314 \\ 
LOSS-CSP2 & -0.044&0.012&0.034&8 & 0.0030 & 0.0213&0.903 \\ 
\hline
2007le    & $V_{\rm stars}$& & & & $V_{\rm SN}$\\
CSP2-CfA4 &  0.042&0.006&0.017&9 & 0.0527 & 0.0201&10.254 \\ 
LOSS-CfA4 & -0.032&0.052&0.074&2 & 0.0258 & 0.0385&2.814 \\ 
LOSS-CSP2 & -0.039&0.036&0.072&4 &-0.0343 & 0.0464&6.137 \\ 
\hline
2008C     & $V_{\rm stars}$& & & & $V_{\rm SN}$\\
CSP2-CfA4 &  0.028&0.004&0.014&15& 0.0227 & 0.0771&4.184 \\ 
LOSS-CfA4 &  0.017&0.004&0.021&23& 0.0295 & 0.0303&1.184 \\ 
LOSS-CSP2 & -0.007&0.005&0.018&11& 0.0021 & 0.0218&1.410 \\ 
\hline
2009dc    & $V_{\rm stars}$& & & & $V_{\rm SN}$\\
CSP2-CfA4 &  0.011&0.007&0.021&10& 0.0221 & 0.0168&1.979 \\ 
LOSS-CfA4 &  0.013&0.007&0.014&4 & 0.0497 & 0.0578&3.230 \\ 
LOSS-CSP2 &  0.017&0.006&0.017&7 & 0.0369 & 0.0191&3.767 \\ 
\enddata
\label{table_3commonsn}
\end{deluxetable}



\clearpage

\clearpage

\clearpage

\clearpage

\clearpage

\clearpage

\clearpage

\clearpage

\clearpage

\clearpage

\clearpage

\clearpage

\end{document}